\documentclass[runningheads]{llncs}
\usepackage[T1]{fontenc}
\usepackage{graphicx}
\usepackage[utf8]{inputenc}

\usepackage{booktabs}
\usepackage{colortbl}
\definecolor{grayrow}{rgb}{.9,.9,.9}
\usepackage{tabularx}

\usepackage{algorithmic}
\usepackage{amsmath}
\usepackage{textcomp}
\usepackage[utf8]{inputenc}
\usepackage{multirow}
\usepackage[table,xcdraw]{xcolor}
\usepackage{todonotes}
\usepackage{rotating}
\usepackage{hyphenat}
\usepackage{varwidth}
\usepackage{makecell}
\usepackage{colortbl}
\usepackage{float}
\usepackage{hhline}
\usepackage{subfigure}
\usepackage{caption}
\usepackage{booktabs}
\usepackage{siunitx}
\usepackage{tcolorbox}
\usepackage{url}
\usepackage{xcolor}
\usepackage{epsfig}
\usepackage{calc}
\usepackage{multicol}
\usepackage[bottom]{footmisc}
\usepackage{mdframed}
\usepackage{graphicx}
\usepackage{hyperref}

\usepackage{academicons}
\usepackage{xcolor}
\definecolor{orcidlogocol}{HTML}{A6CE39} 

\definecolor{bggray}{RGB}{242,242,242}

\mdfsetup{
  backgroundcolor=bggray,
  linewidth=1pt,
  linecolor=black,
}

\begin{document}
\title{Unifying Economic and Language Models for Enhanced Sentiment Analysis of the Oil Market
	\thanks{This is a pre-print version of the paper. The final version has been published in the Lecture Notes in Business Information Processing (LNBIP,volume 518) and is available on \url{https://link.springer.com/chapter/10.1007/978-3-031-64748-2_6}}
}
\titlerunning{Unifying Economic and Language Models for Enhanced Sentiment Analysis}
%

\author{
    Himmet Kaplan\inst{1} \href{https://orcid.org/0000-0002-1115-8669}{\includegraphics[scale=0.06]{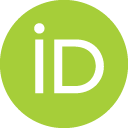}} \and 
    Ralf-Peter Mundani\inst{2} \href{https://orcid.org/0000-0001-6248-714X}{\includegraphics[scale=0.06]{orcid.png}} \and 
    Heiko Rölke\inst{2} \href{https://orcid.org/0000-0002-9141-0886}{\includegraphics[scale=0.06]{orcid.png}} \and 
    Albert Weichselbraun\inst{2} \href{https://orcid.org/0000-0001-6399-045X}{\includegraphics[scale=0.06]{orcid.png}} \and 
    Martin Tschudy\inst{2} \href{https://orcid.org/0009-0009-4093-387X}{\includegraphics[scale=0.06]{orcid.png}}
}
\authorrunning{Kaplan et al.}
\institute{
    Zurich University of Applied Sciences, Winterthur 8401, Switzerland \\
    \email{kapa@zhaw.ch} \and
    University of Applied Sciences of the Grisons, Chur 7000, Switzerland \\
    \email{\{firstname.lastname\}@fhgr.ch}
}

%
\maketitle              
\begin{abstract}
Crude oil, a critical component of the global economy, has its prices influenced by various factors such as economic trends, political events, and natural disasters. Traditional prediction methods based on historical data have their limits in forecasting, but recent advancements in natural language processing bring new possibilities for event-based analysis. In particular, Language Models (LM) and their advancement, the Generative Pre-trained Transformer (GPT), have shown potential in classifying vast amounts of natural language. However, these LMs often have difficulty with domain-specific terminology, limiting their effectiveness in the crude oil sector. Addressing this gap, we introduce CrudeBERT, a fine-tuned LM specifically for the crude oil market. The results indicate that CrudeBERT's sentiment scores align more closely with the WTI Futures curve and significantly enhance price predictions, underscoring the crucial role of integrating economic principles into LMs.

\keywords{BERT  \and Crude Oil Market \and Fine-Tuning \and GPT \and Language Models \and Large Language Models \and Sentiment Analysis \and Transformers.}
\end{abstract}
\section{Introduction}
\label{sec:introduction}
Crude oil plays a key role as both a primary energy source and raw material and serves as a fundamental indicator for the global economic landscape, whether in a boom or recession. Given the finite characteristics of crude oil as a natural resource, its price is expected to be influenced by its availability which is generally driven by the dynamics of supply and demand. 
However, the literature notes that the availability of crude oil is exposed to highly volatile factors such as economic cycles, geopolitical disturbances, and natural disasters~\cite{buyuksahin_speculators_2011}. 
To address this uncertainty, decision-makers and analysts traditionally utilized technical analysis of structured market data. Yet, this approach has serious limitations due to its dependence on historical data, which often fail to offer significant predictive insights~\cite{mccarthy_applying_2019}.
Empirical studies have indicated that augmenting technical analysis with timely event-based information, such as news, can significantly enhance the reliability of predicting substantial market shifts in publicly traded assets, such as stocks~\cite{qian_stock_2007}. 
As a result, academic research that focused on assessing the effectiveness of news data integration for predictive insights~\cite{baboshkin_multi-source_2021} has gained increasing traction. Wex et al., for instance, indicate that sentiment scores were a statistically significant feature in forecasting models~\cite{wex_early_2013}.
However, interpreting an extensive volume of incoming news data cost-effectively poses natural challenges due to the predominantly textual and, thus, unstructured nature of such content. 
Sentiment analysis has emerged as a promising technique for addressing these challenges, particularly given observations by~\cite{jiang_smart_2020} which indicate that modern sentiment classifiers can achieve remarkable accuracies of up to 97.5\,\%. 

The presented research draws upon FinBERT, a Bidirectional Encoder Representation from Transformers (BERT), a Language Model (LM) for financial sentiment analysis that has been pre-trained for the general financial market. Furthermore, we harness the power of a large Generative Pre-trained Transformer (GPT). In addition, this paper presents improved and extended findings of~\cite{kaplan_conference_2023} which was published at the 25th International Conference on Enterprise Information Systems in which the main contributions can be summarized as follows: (i) developing a method that equips LMs with the capability to identify the major supply and demand factors that drive crude oil futures markets; (ii) evaluating a financial LM named FinBERT by incorporating the economic model of supply and demand into these models and introducing CrudeBERT; (iii) conducting extensive experiments that draw upon multiple prediction settings to benchmark the developed method against a baseline (random binary classification) and two state-of-the-art sentiment analysis frameworks. 

In addition, this extended and revised version (iv) assesses the capabilities of a Large Language Model (LLM), specifically GPT 3.5 from OpenAI through various simulation scenarios to identify an optimal prompt for classifying news regarding changes in the availability of crude oil; and (v) conducts experiments to measure the classification performance of the three sentiment analysis frameworks against a silver benchmark followed by a quantitative evaluation that contrasts the best-performing model for prediction of the following day prices.
 
\section{Related Work}
\label{sec:related-work}
Crude oil greatly impacts the global economy, making its availability and supply chain a topic of extensive research and analysis. Numerous scholarly articles have been written on the subject, exploring various analytical methods, including technical analysis and fundamental assessments, to forecast crude oil prices. This literature review focuses on studies incorporating sentiment-related factors of news to estimate changes in supply and demand to ultimately predict crude oil prices. To achieve this goal, this chapter first examines the Efficient Market Hypothesis. Afterward, it discusses the role of sentiment analysis in finance, tracing its development through various techniques. Furthermore, it explores the use of traditional and state-of-the-art Natural Language Processing (NLP) techniques, covering conventional lexicon-based approaches to the modern transformer-era methodologies that have given rise to the prominent LLMs of today.

\subsection{Market Efficiency and the Role of Sentiment Analysis}
\label{sec:market-efficiency}
The Efficient Market Hypothesis (EMH) is a foundational concept in finance that investigates the influence of both public and non-public information on the predictability of financial markets. Eugene Fama categorizes EMH into three forms: weak, semi-strong, and strong~\cite{fama_efficient_1970}. The weak form believes prices are driven exclusively by historical data, discounting the effect of external information sources such as news articles or social media. In contrast, the semi-strong form incorporates both historical prices and public data, suggesting that only non-public insights like insider details can allow forecasting~\cite{malkiel_efficient_1989}. The strong form encompasses historical, public, and confidential data, arguing that any information-based analysis cannot yield consistently higher returns. 

However, studies such as that by Qian and Rasheed demonstrate the viability of technical analysis of price fluctuations which were able to deliver predictive accuracies of over 50\,\%~\cite{qian_stock_2007}. Similarly, the findings by Gu indicate that relying on historical price-based approaches with deep learning may surpass the performance of sentiment-based indicators~\cite{gu2020prediction}. 
However, Hu's study points out the limited presence of modern sentiment-based analysis in leading journals~\cite{hu2021survey}. 
The significance of news media grew substantially during the pandemic, as demonstrated in a study analyzing social media to grasp shifts in public behavior~\cite{liu2021monitoring} including sentiment analysis of news media~\cite{balaji2017survey}. This perspective aligns with the findings of Mahata et al., especially after the early 2020 pandemic market crash, about the potential inefficacy of existing models based on historical data in the current markets~\cite{mahata2021characteristics}. During this period, the pandemic's effects and the subsequent vaccine releases caused big changes in consumer sentiment and stock prices.
Thus, contemporary price prediction models aim to adopt various methods including news analytics~\cite{rousidis2020social}. In terms of structuring news to be used with other tabular data, sentiment analysis is considered a prevalent classification task. It aims to categorize affective and subjective information within entire documents, paragraphs, and sentences.

\subsection{Evolution of Sentiment Analysis in Finance}
\label{sec:evolution-nlp}
Effectively interpreting the sentiment of news data streams is challenging due to their textual and unstructured character. While sentiment can encompass a wide range of emotions, financial literature often simplifies it into binary polarities to directly analyze market trends~\cite{li_news_2014}. One of the earliest methods for financial sentiment analysis (FSA) in natural language was the Bag-of-Words (BOW) approach developed in the 1980s, which is also known as the lexicon-based method~\cite{liew_fine-grained_nodate}. In this methodology, the overall sentiment of a text is determined by summing up the sentiment scores of both positive and negative words. As implied by its title, BOW methods employ a lexicon with words and their assigned sentiment values, ideally curated by several human evaluators. 

Notably, a prominent lexicon for FSA was developed by Loughran and McDonald, designed to interpret liabilities related to 10-K filing returns~\cite{loughran_when_2011}. In their subsequent research~\cite{loughran2016textual}, they presented a study concerning text analysis, specifically in the areas of accounting and finance. However, while these lexicons offer a structured approach to sentiment analysis, the flexible nature of language necessitates continuous adaptation and refinement. In particular, creating comprehensive lexicons that capture all potential keywords and their combinations is a difficult task, especially because word sentiment can change depending on the surrounding context.

Parallel to these developments, rapid advances in computing capabilities paved the way for the increasing influence of machine learning in FSA methodologies. The study by~\cite{cui2023survey} highlights that sentiment analysis research has experienced dramatic growth from just two publications in 2002 to 1466 in 2021.

Before the era of transformer models (Section \ref{sec:evolution-transformers}), supervised machine learning emerged as a prominent method for FSA. This approach has gained traction primarily due to its ability to leverage human-curated training datasets, especially for tasks such as classification~\cite{chollet_deep_2018}. Machine learning algorithms, under the umbrella of supervised learning, improve their performance by learning from labeled training data. Through this process, they discern complex rules and patterns, with the ultimate goal of achieving generalizations.

For instance, Recurrent Neural Networks (RNN) and its variant, Long Short-Term Memory (LSTM), have been tailored towards analyzing sequential data types such as text~\cite{tang-etal-2016-effective}. However, the need for large training datasets can be seen as a major limitation, particularly in sentiment classification tasks where high precision is required. Creating the necessary training datasets requires significant input from domain experts. A notable challenge with RNNs is their vulnerability to vanishing and exploding gradients, combined with their incompatibility with parallel processing, leading to extended training durations~\cite{chollet_deep_2018}.

Combining supervised techniques with unsupervised machine learning models mitigates some of these problems. In this realm, word embeddings or word vector models, which represent words in a vector space based on semantic similarity, have gained prominence. Well-known models like word2vec~\cite{mikolov_2013} and Glove~\cite{pennington_glove_2014} can be trained on large text corpora, capturing intricate word semantics. Yet, their inability to distinguish context remains an important limitation. Therefore, word embeddings assign identical vectors to words irrespective of their context.

\subsection{The Advent of the Attention Mechanism and Transformers}
\label{sec:evolution-transformers}
\begin{figure}
\includegraphics[width=1\textwidth]{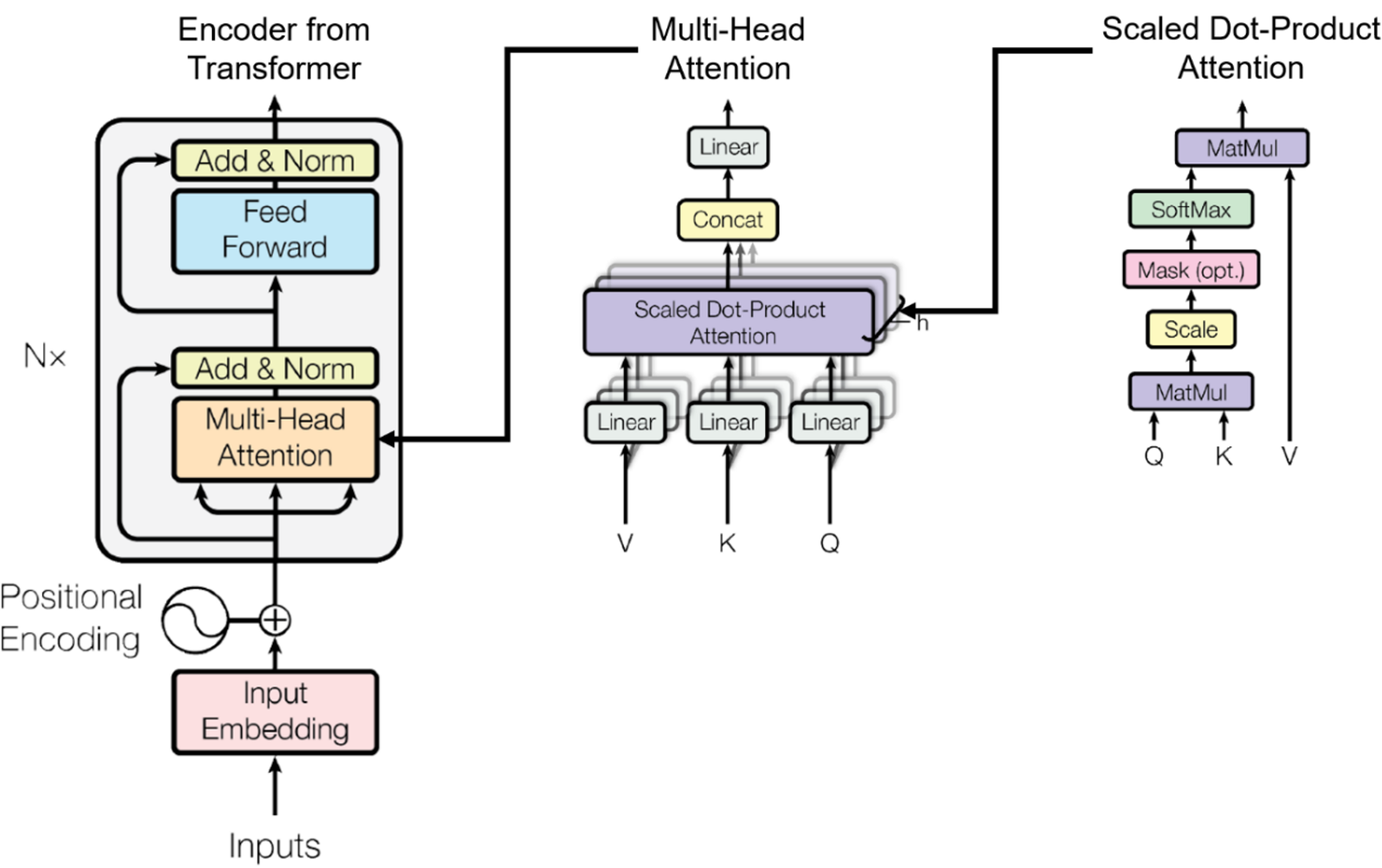}
\caption{Components of the multi-head attention mechanism~\cite{vaswani_attention_2017}.} 
\label{fig:components-attention}
\end{figure}

Advancements in neural network architectures have been instrumental in shaping the trajectory of contextualized word embeddings, also known as language models (LMs). At the core of these developments lies the attention mechanism, a pivotal technique that facilitates neural networks in allocating computational resources effectively by focusing attention on significant data~\cite{bahdanau_neural_2016}. Building upon this mechanism, Vaswani et al.~\cite{vaswani_attention_2017}introduced the transformer architecture. Designed for enhanced parallel processing, transformers extensively leverage the attention mechanism. Initially conceptualized for neural machine translation, the transformer integrates an encoder consisting of a series of multi-headed attention layers, allowing the model to analyze sequences from a multifaceted perspective, as depicted in Figure~\ref{fig:components-attention}. The capability to capture the context of words together with the option to draw upon and customize large pre-trained models has been key to the success of transformer-based for many further NLP applications, including sentiment analysis.

\subsection{Language Models for Sentiment Analysis}
\label{sec:bert}
Following the introduction of the transformer architecture, Devlin et al.~\cite{devlin_bert_2018} recognized the encoder's potential as a robust representation learning mechanism. This realization gave rise to Bidirectional Encoder Representations from Transformers (BERT), one of the first LMs, which has demonstrated high adaptability across diverse NLP tasks and an unparalleled ability for contextual interpretation of words~\cite{yenicelik_understanding_2020}. BERT was developed in two configurations: the smaller version, BERT-base, with 110 million parameters, and its larger counterpart, BERT-large, with 340 million parameters~\cite{devlin_bert_2018}. 

\begin{figure}
	\includegraphics[width=\textwidth]{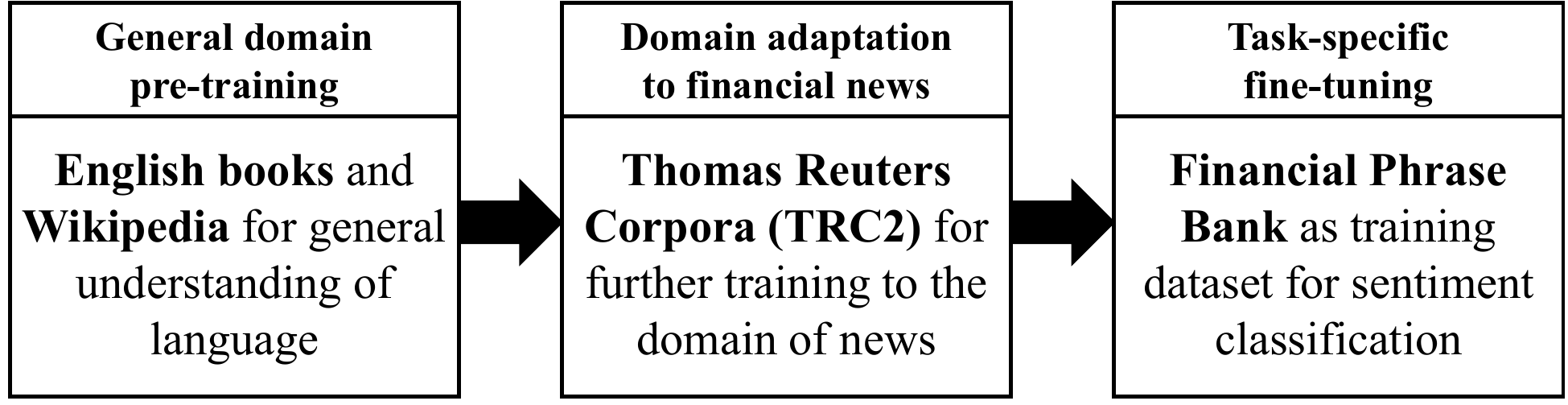}
\caption{Process of generating FinBERT~\cite{kaplan_conference_2023}.} 
\label{fig:finbert-process}
\end{figure}

Although initial training for LMs was predominantly based on generalized resources such as English \textit{Wikipedia} and the \textit{BookCorpus}, it became evident that specialized understanding was necessary in certain sectors. Driven by this need Araci et al. developed FinBERT~\cite{araci_finbert_2019}, which is trained on the Thomson Reuters Text Research Collection (TRC2) and further refined using the \textit{FinancialPhraseBank} dataset~\cite{malo_good_2014} (Figure~\ref{fig:finbert-process}). Despite its capabilities, the application of FinBERT in niche domains, such as the crude oil market is relatively sparse.

\subsection{The Emergence of Large Language Models}
\label{sec:llm}
With the solid foundation established by transformers and their offspring LMs, research interests naturally gravitated towards developing models of even larger scale and complexity. Manifesting this trend is GPT-3 which comprises 175 billion parameters~\cite{brown_language_2020}. In contrast to models such as BERT, GPT-3 is designed without an encoder, focusing on a decoder-centric architecture that excels in text generation tasks, ranging from translation to summarization. Although GPT-3's design is not inherently optimized for classification tasks such as sentiment analysis, its extensive training dataset and proficiency in natural language understanding have cemented its position as a leading model in language processing.

\section{Methodology}
\label{sec:method}
This section outlines the methods used for computing text sentiment, and for evaluating their suitability for predicting changes in crude oil prices. A key element of this research is the comparative assessment between FinBERT, GPT-3.5, and a proprietary solution, the RavenPack Event Sentiment Score, in their ability to classify news articles based on their expected impact on crude oil prices. The subsequent sections discuss the datasets used, preprocessing techniques, challenges faced during data preprocessing, and a detailed introduction to each sentiment classifier.

\subsection{Datasets and Data Sources}
\label{sec:method-datasets}

\subsubsection{News Dataset}
\label{sec:method-news-data}
The news dataset, sourced from the \textit{RavenPack Realtime News Discovery} platform, comprises about 46,000 headlines covering the period from 1 January 2000 to 1 April 2021. These headlines are closely related to the domain of crude oil. The sourced headlines originate from around 950 distinct news sources. Notably, about half of these headlines originate from prominent news agencies such as \textit{Dow Jones}, \textit{Reuters}, \textit{Bloomberg News}, and \textit{Platts}. However, to capture a more exhaustive and representative coverage of the news landscape, evaluations prioritized the period post-2012. This decision was influenced by the expanded availability of diverse sources in the latter years, resulting in approximately 26,500 headlines from 1 January 2012 to 1 April 2021. The focus on headlines aligns with the research of Li et al., where headlines were used for their accessibility and computational efficiency since they are assumed to serve as a summary of the full article~\cite{li_text-based_2019}. 

\subsubsection{Price Dataset}
\label{sec:method-price-data}
The Brent Crude and Western Texas Intermediate (WTI) standards are popular crude oil price indices. Due to the predominance of the English language in the news dataset, WTI futures prices, notably relevant in the U.S. context, were chosen for this research. The historical price datasets were obtained from the \textit{investing.com} platform, matching the timeline of the news headlines.

\subsection{Sentiment Classifiers}
\label{sec:sentiment-classifiers}
\subsubsection{RavenPack Event Sentiment Score}
RavenPack, a renowned analytics firm, developed the \textit{Event Sentiment Score} (ESS). It's a sophisticated index that uses a vast amount of lexica, annotated by experts, to measure the sentiment of news. The ESS can vary between -1 (negative sentiment) and 1 (positive sentiment) to capture the mood of news. The ESS is designed to recognize various sentiment signals found in financial news. It can assess the sentiment of different events, from company earnings to natural disasters, tailored to specific assets~\cite{hafez_ess_factor_2020}. For our study, the ESS was chosen as an additional benchmark for comparison.

\subsubsection{FinBERT}
\label{sec:finbert}
FinBERT was built upon the BERT model and is one of the pioneering language models for financial news. Unlike social media, news media are less likely to contain sarcasm, grammatical errors, or slang. Thus, FinBERT adapts the domain of BERT by performing additional training on a portion of Reuters’ TRC2, containing 1.8M news articles published between 2008 and 2010. For sentiment classification, FinBERT was trained on the Financial PhraseBank, which consists of 4845 English sentences randomly selected from the LexisNexis financial news database. These sentences were then annotated by 16 experts from the finance and business fields. When given an input, it provides a granular sentiment score ranging from -1 (negative sentiment) to 1 (positive sentiment).

\subsubsection{GPT-3.5}
\label{sec:gpt}
Within the realm of LLMs, GPT-3.5 stands as a state-of-the-art language model, renowned for its diverse application capabilities. It excels in generating coherent and contextually relevant text across various domains. In this study, we used the most recent iteration from the GPT-3.5 series, text-davinci-003, accessed via the API~\cite{openai_url}. The following input prompt was employed for sentiment classification:

\noindent \textit{"Classify the sentiment of the following headlines as either 'Positive', 'Negative', or 'Neutral'. Return only the ID and your classification as a dictionary."}

\noindent The model produces a discrete score of -1 (negative sentiment), 0 (neutral sentiment), and 1 (positive sentiment). The effectiveness and optimization of this prompt are discussed in Section~\ref{sec:evaluation-gpt}.

\subsection{Data Preprocessing and Normalization}
\label{sec:method-data-preprocessing}
To handle inconsistencies in the dataset due to absent headlines or missing prices caused by market closures on the weekends and holidays, rows with incomplete data were discarded.

\subsubsection{Sentiment Normalization}
Sentiments extracted from the headlines were normalized using z-statistics. By normalizing sentiment data over a sliding window we account for the EMH (i.e., the market price reflects all publicly available information) by assuming that only new information that causes a change in expectations results in significant price changes. This normalization was applied over a sliding window for a broad weekly sentiment perspective taking into account market changes and newly arriving information:
\begin{equation}
\label{eq:sent_norm}
sent_{norm, t} = \frac{sent_{t}-\overline{sent}_{t,w}}{\sigma_{t,w}} 
\end{equation}

\noindent Here, \(\overline{sent}_{t, w}\) represents the mean sentiment over a moving time window starting from the current time point \( t \) and going back \( w \) time points to \( t-w \), and \(\sigma_{t,w}\) denotes the standard deviation within the same period. Furthermore, the output of the sentiment classifiers had to be normalized. For instance, granular sentiment scores from RavenPack and FinBERT typically provide scores in the continuous range, signifying the intensity of the sentiment. For instance, a score of 0.2 might indicate a slightly positive sentiment, while a score of -0.8 could imply a strongly negative sentiment. For compatibility with the values obtained from GPT 3.5, which were not in float format but rather categorical (1 for positive, 0 for neutral, and -1 for negative), it was necessary to remap the granular sentiment scores \( x \) based on the observed distribution into these discrete categories as follows:

\begin{equation*}
\begin{array}{lcl}
\phantom{-}0.1 < x \leq \phantom{-}1.0 & \mapsto & \phantom{-}1 \\  
-0.1 \leq x \leq \phantom{-}0.1 & \mapsto & \phantom{-}0 \\ 
-1.0 \leq x < -0.1 & \mapsto & -1
\end{array}
\end{equation*}

\subsubsection{Price Normalization} 
The works of Hamilton indicated that the oil price appears to be influenced by a random walk with drift due to market volatility and, therefore, can show random fluctuations that overlap short-term and long-term trends within the market~\cite{hamilton_understanding_2008}. Thus, using a similar method as sentiment normalization, the price data was processed also using z-statistics to better distinguish between significant market movements and random fluctuations:

\begin{equation}
price_{norm,t} = \frac{price_t-\overline{price}_{t,w}}{\sigma_{t,w}} 
\end{equation}
In this equation, $\overline{price}_{t,w}$ indicates the average price, and $\sigma_{t,w}$ represents the standard deviation over the chosen moving time window.

\noindent One of the key goals in evaluating predictive performance was to compare daily sentiment scores with price returns on the following day. This alignment was based on the assumption that market reactions to news typically occur within a 24-hour window. For this synchronization, daily $Returns$ for WTI futures were calculated as the basis for comparing sentiment scores to market reactions:
\begin{equation}
\label{eq:return}
Return = \frac{Price_t-{Price}_{t-1}}{{Price}_{t-1}} 
\end{equation}

\section{Evaluation}
\label{sec:evaluation}
This section delves into the comparative performance analysis of different LMs which include fine-tuning LMs in terms of the expectations when economic theories are considered, leading to the creation of CrudeBERT. In addition,  evaluation further extends into the exploration of various prompt optimizations of an LLM, providing an understanding of how different contexts and examples influence its sentiment classification performance. Lastly, an evaluation of the sentiment classifiers, focusing on their ability to predict the subsequent day's crude oil futures was conducted. Specifically, classifiers such as FinBERT, RavenPack ESS, GPT 3.5, and the newly developed CrudeBERT within this chapter. 

\subsection{Interplay of Supply and Demand}
\label{sec:supply-demand}
An initial exploration involved manual annotation of sample headlines based on their expected impact on crude oil supply and demand. The notion behind this approach was the economic model introduced by Adam Smith in 1776~\cite{smith_inquiry_1776}. He argued that the value of commodities, like crude oil, is fundamentally influenced by their supply and demand. In this scenario, \textit{supply} signifies the volume of a commodity available for sale at a specific price during a particular time frame. Conversely, \textit{demand} signifies the volume that buyers seek at that price during that period. The convergence of these two forces shapes a competitive landscape where the price finds balance through their equilibrium. This equilibrium can be visualized with scenarios such as:

\begin{equation*}
\begin{array}{llclcl}
\text{Less supply} & \text{\& same demand} & \mapsto & \text{shortage} & \mapsto & \text{higher prices.} \\ 
\text{More supply} & \text{\& same demand} & \mapsto & \text{surplus} & \mapsto & \text{lower prices.} \\ 
\text{Less demand} & \text{\& same supply} & \mapsto & \text{surplus} & \mapsto & \text{lower prices.} \\ 
\text{More demand} & \text{\& same supply} & \mapsto & \text{shortage} & \mapsto & \text{higher prices.}
\end{array}
\end{equation*}

\noindent Given this foundational understanding, our expectations for sentiment classification of news sentiment consider shortage as \textit{Positive}, \textit{Neutral} for stagnant scenarios, and \textit{Negative} for news indicating a surplus in crude oil availability.

\subsection{Preliminary Analysis}
Initial observations, presented in Table~\ref{tab:should-sentiment}, suggest that the sentiment classifications of FinBERT and GPT 3.5 deviate from our expectations outlined in Section~\ref{sec:supply-demand}. These disparities are particularly noteworthy, given that crude oil is a publicly traded commodity, and FinBERT's training is grounded in the wider context of financial news. This unexpected deviation instigates a deeper investigation to understand the underlying causes.

\begin{table}
\captionof{table}{Labeled headlines with sentiment classifications of an LM and LLM.}
\label{tab:should-sentiment}
\centering
\resizebox{\columnwidth}{!}{%
\begin{tabular}{|c|c|c|c|c|} 
\hline
\multicolumn{2}{|c|}{\textbf{Headlines}}                                                                                                                                                                                         & \begin{tabular}[c]{@{}c@{}}\textbf{Sentiment }\\\textbf{Expected}\end{tabular} & \begin{tabular}[c]{@{}c@{}}\textbf{Sentiment }\\\textbf{FinBERT}\end{tabular}  & \begin{tabular}[c]{@{}c@{}}\textbf{Sentiment }\\\textbf{GPT 3.5}\end{tabular} \\ 
\hline
\multirow{6}{*}{\makecell{\textbf{Shortage}}} & \begin{tabular}[c]{@{}c@{}}\textbf{Major Explosion, Fire at Oil }\\\textbf{Refinery in Southeast Philadelphia}\end{tabular} & {\cellcolor[rgb]{0.388,0.745,0.482}}Positive  & {\cellcolor[rgb]{0.973,0.412,0.42}}Negative & {\cellcolor[rgb]{0.973,0.412,0.42}}Negative \\ 
\hhline{|~----|}
                                               & \begin{tabular}[c]{@{}c@{}}\textbf{PETROLEOS confirms Gulf of }\\\textbf{Mexico oil platform accident}\end{tabular}         & {\cellcolor[rgb]{0.388,0.745,0.482}}Positive  & {\cellcolor[rgb]{0.973,0.412,0.42}}Negative & {\cellcolor[rgb]{0.973,0.412,0.42}}Negative \\ 
\hhline{|~----|}
                                               & \begin{tabular}[c]{@{}c@{}}\textbf{CASUALTIES FEARED AT OIL }\\\textbf{ACCIDENT NEAR IRANS BORDER }\end{tabular}            & {\cellcolor[rgb]{0.388,0.745,0.482}}Positive  & {\cellcolor[rgb]{0.973,0.412,0.42}}Negative & {\cellcolor[rgb]{0.973,0.412,0.42}}Negative \\ 
\hhline{|~----|}
                                               & \begin{tabular}[c]{@{}c@{}}\textbf{EIA Chief expects Global Oil }\\\textbf{Demand Growth 1 M B/D to 2011}\end{tabular}      & {\cellcolor[rgb]{0.388,0.745,0.482}}Positive  & {\cellcolor[rgb]{0.388,0.745,0.482}}Positive  & {\cellcolor[rgb]{0.388,0.745,0.482}}Positive \\ 
\hhline{|~----|}
                                               & \begin{tabular}[c]{@{}c@{}}\textbf{Turkey Jan-Oct Crude }\\\textbf{Imports +98.5\% To 57.9M MT}\end{tabular}                & {\cellcolor[rgb]{0.388,0.745,0.482}}Positive  & {\cellcolor[rgb]{0.388,0.745,0.482}}Positive  & {\cellcolor[rgb]{0.388,0.745,0.482}}Positive \\ 
\hhline{|~----|}
                                               & \begin{tabular}[c]{@{}c@{}}\textbf{China's crude oil imports }\\\textbf{up 78.30\% in February 2019}\end{tabular}           & {\cellcolor[rgb]{0.388,0.745,0.482}}Positive  & {\cellcolor[rgb]{0.388,0.745,0.482}}Positive  & {\cellcolor[rgb]{0.388,0.745,0.482}}Positive \\ 
\hline
\multirow{6}{*}{\makecell{\textbf{Stagnant}}} & \begin{tabular}[c]{@{}c@{}}\textbf{Russia Energy Agency: Sees }\\\textbf{Oil Output Flat In 2005}\end{tabular}               & {\cellcolor[rgb]{0.996,0.922,0.518}}Neutral   & {\cellcolor[rgb]{0.973,0.412,0.42}}Negative  & {\cellcolor[rgb]{0.996,0.922,0.518}}Neutral \\ 
\hhline{|~----|}
                                               & \begin{tabular}[c]{@{}c@{}}\textbf{Malaysia Oil Production Steady }\\\textbf{This Year At 700,000 B/D}\end{tabular}          & {\cellcolor[rgb]{0.996,0.922,0.518}}Neutral   & {\cellcolor[rgb]{0.388,0.745,0.482}}Positive  & {\cellcolor[rgb]{0.996,0.922,0.518}}Neutral \\ 
\hhline{|~----|}
                                               & \begin{tabular}[c]{@{}c@{}}\textbf{ExxonMobil:Nigerian Oil Output }\\\textbf{Unaffected By Union Threat}\end{tabular}        & {\cellcolor[rgb]{0.996,0.922,0.518}}Neutral   & {\cellcolor[rgb]{0.973,0.412,0.42}}Negative  & {\cellcolor[rgb]{0.996,0.922,0.518}}Neutral \\ 
\hhline{|~----|}
                                               & \begin{tabular}[c]{@{}c@{}}\textbf{Yukos July Oil Output Flat }\\\textbf{On Mo, 1.73M B/D - Prime-Tass}\end{tabular}          & {\cellcolor[rgb]{0.996,0.922,0.518}}Neutral   & {\cellcolor[rgb]{0.973,0.412,0.42}}Negative  & {\cellcolor[rgb]{0.996,0.922,0.518}}Neutral \\ 
\hhline{|~----|}
                                               & \begin{tabular}[c]{@{}c@{}}\textbf{2nd UPDATE: Mexico's Oil Output }\\\textbf{Unaffected By Hurricane}\end{tabular}           & {\cellcolor[rgb]{0.996,0.922,0.518}}Neutral    & {\cellcolor[rgb]{0.973,0.412,0.42}}Negative  & {\cellcolor[rgb]{0.996,0.922,0.518}}Neutral \\ 
\hhline{|~----|}
                                               & \begin{tabular}[c]{@{}c@{}}\textbf{UPDATE: Ecuador July Oil }\\\textbf{Exports Flat On Mo At 337,000 B/D}\end{tabular}       & {\cellcolor[rgb]{0.996,0.922,0.518}}Neutral   & {\cellcolor[rgb]{0.996,0.922,0.518}}Neutral  & {\cellcolor[rgb]{0.996,0.922,0.518}}Neutral \\ 
\hline
\multirow{6}{*}{\makecell{\textbf{Surplus}}} & \begin{tabular}[c]{@{}c@{}}\textbf{China February Crude }\\\textbf{Imports -16.0\% On Year}\end{tabular}               & {\cellcolor[rgb]{0.973,0.412,0.42}}Negative   & {\cellcolor[rgb]{0.388,0.745,0.482}}Positive  & {\cellcolor[rgb]{0.973,0.412,0.42}}Negative \\ 
\hhline{|~----|}
                                               & \begin{tabular}[c]{@{}c@{}}\textbf{Turkey May Crude Imports }\\\textbf{down 11.0\% On Year}\end{tabular}        & {\cellcolor[rgb]{0.973,0.412,0.42}}Negative   & {\cellcolor[rgb]{0.973,0.412,0.42}}Negative  & {\cellcolor[rgb]{0.973,0.412,0.42}}Negative \\ 
\hhline{|~----|}
                                               & \begin{tabular}[c]{@{}c@{}}\textbf{Japan June Crude Oil Imports }\\\textbf{decrease 10.9\% On Yr}\end{tabular}            & {\cellcolor[rgb]{0.973,0.412,0.42}}Negative   & {\cellcolor[rgb]{0.973,0.412,0.42}}Negative  & {\cellcolor[rgb]{0.973,0.412,0.42}}Negative \\ 
\hhline{|~----|}
                                               & \begin{tabular}[c]{@{}c@{}}\textbf{Iran’s’ Feb Oil Exports +20.9\% }\\\textbf{On Mo at 1.56M B/D - Official}\end{tabular}      & {\cellcolor[rgb]{0.973,0.412,0.42}}Negative   & {\cellcolor[rgb]{0.996,0.922,0.518}}Neutral  & {\cellcolor[rgb]{0.388,0.745,0.482}}Positive \\ 
\hhline{|~----|}
                                               & \begin{tabular}[c]{@{}c@{}}\textbf{Apache announces large petroleum }\\\textbf{discovery in Philadelphia}\end{tabular}                & {\cellcolor[rgb]{0.973,0.412,0.42}}Negative   & {\cellcolor[rgb]{0.996,0.922,0.518}}Neutral  & {\cellcolor[rgb]{0.388,0.745,0.482}}Positive \\ 
\hhline{|~----|}
                                               & \begin{tabular}[c]{@{}c@{}}\textbf{Turkey finds oil near }\\\textbf{Syria, Iraq border}\end{tabular}                 & {\cellcolor[rgb]{0.973,0.412,0.42}}Negative   & {\cellcolor[rgb]{0.996,0.922,0.518}}Neutral  & {\cellcolor[rgb]{0.388,0.745,0.482}}Positive \\ 
\hline
\end{tabular}%
}
\end{table}

\noindent In situations where there's a decrease in supply (perhaps due to unforeseen incidents at oil infrastructures), both models lean toward negative sentiment. This is unexpected since such events would typically bolster crude oil prices. One plausible reason for this could be the natural perception that incidents, especially in finance, are considered harmful not just because of the potential operational disruptions but also because of the reputational implications they might carry. A closer look at the \textit{Financial Phrase Bank} supports this for FinBERT. In contrast, signals of surging demand are typically seen as catalysts for price appreciation. Our experiments with both FinBERT and GPT 3.5 align with this hypothesis.

Conversely, FinBERT generally categorizes headlines that indicate no changes in availability as negative. This could be attributed to the financial sector's inclination towards growth, where no change might be perceived as a setback. GPT 3.5, however, seems to be more aligned with the expectations in this context. Furthermore, a decrease in demand should ideally signal an oversupply, leading to price reduction. This interpretation is largely ratified by FinBERT for most headlines, except for a few anomalies. For instance, GPT 3.5 and FinBERT differ in their assessment of a headline indicating a -16.0\,\% in imports, with the latter seemingly struggling with the negative numbers, which was also observed with similar examples. Lastly, GPT-3.5 misclassified topics, notably those related to oil discoveries and imports, even with an optimized prompt. Such inconsistencies indicate the potential limitations of GPT-3.5 in classifying oil-related news.

These observations resonate with findings from Xing et al., suggesting common challenges in using universal sentiment tools for niche sectors. This phenomenon is also known as the domain adaptation problem~\cite{xing_financial_2020}. This has also been highlighted by the research of Leippold~\cite{leippold_sentiment_2023}, emphasizing the unique language intricacies of specialized domains like climate studies. Potential solutions, as proposed by Weichselbraun et al., involve constructing tailored sentiment models for specific sectors~\cite{weichselbraun_automatic_2022}.

\subsection{Integration of Language Models with Economic Principles}
\label{sec:fine-tune}
The low effectiveness of FinBERT, as noted in \ref{tab:should-sentiment}, motivated an attempt to include the principles of supply and demand into FinBERT such that it behaves to the expectations detailed in \ref{sec:supply-demand}. This integration was carried out through domain-specific fine-tuning to crude oil, which led to the development of CrudeBERT.

\subsubsection{Constructing a Domain-Specific Dataset for Crude Oil}
\begin{figure}
\centering
\includegraphics[width=\textwidth]{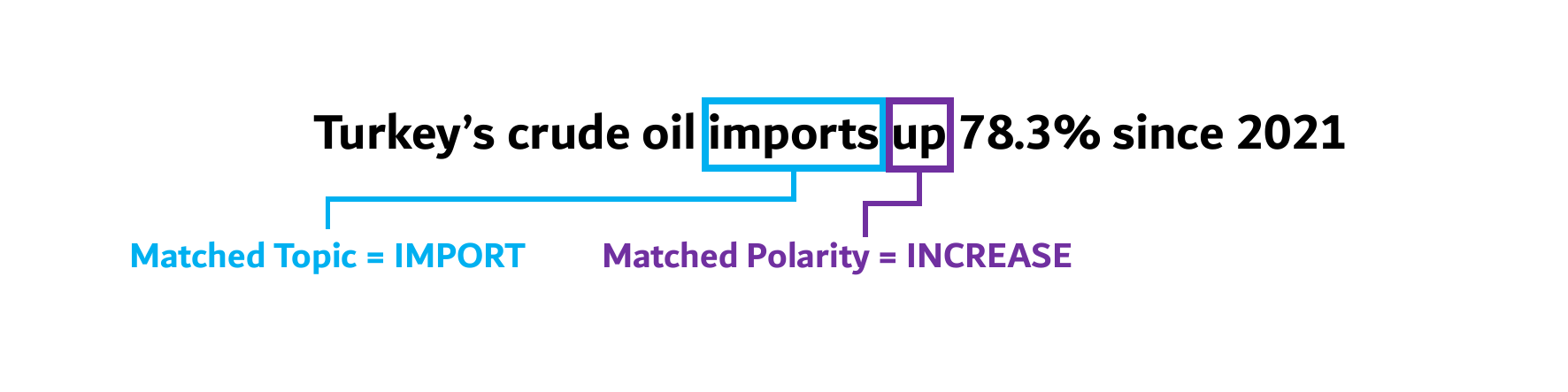}
\caption{Illustration of an identified topic and polarity~\cite{kaplan_conference_2023}}.
\label{fig:identifying-topics}
\end{figure}

To construct a domain-specific training dataset for crude oil, an analysis was conducted on numerous headlines to identify the most frequently recurring topics and the keywords that are representative of these topics. This process identified the following major topics: \textit{accidents, oil discoveries, changes in exports, changes in imports, changes in demand, pricing, supply, pipeline limitations, drilling,} and \textit{spillage}. These identified keywords were then utilized on the news dataset to retrieve relevant headlines and subsequently were categorized under that topic. As an illustration, the headline showcased in Figure \ref{fig:identifying-topics} was aligned with the topic \emph{increased imports}. 

\begin{figure}[H]
\centering
\includegraphics[width=\textwidth]{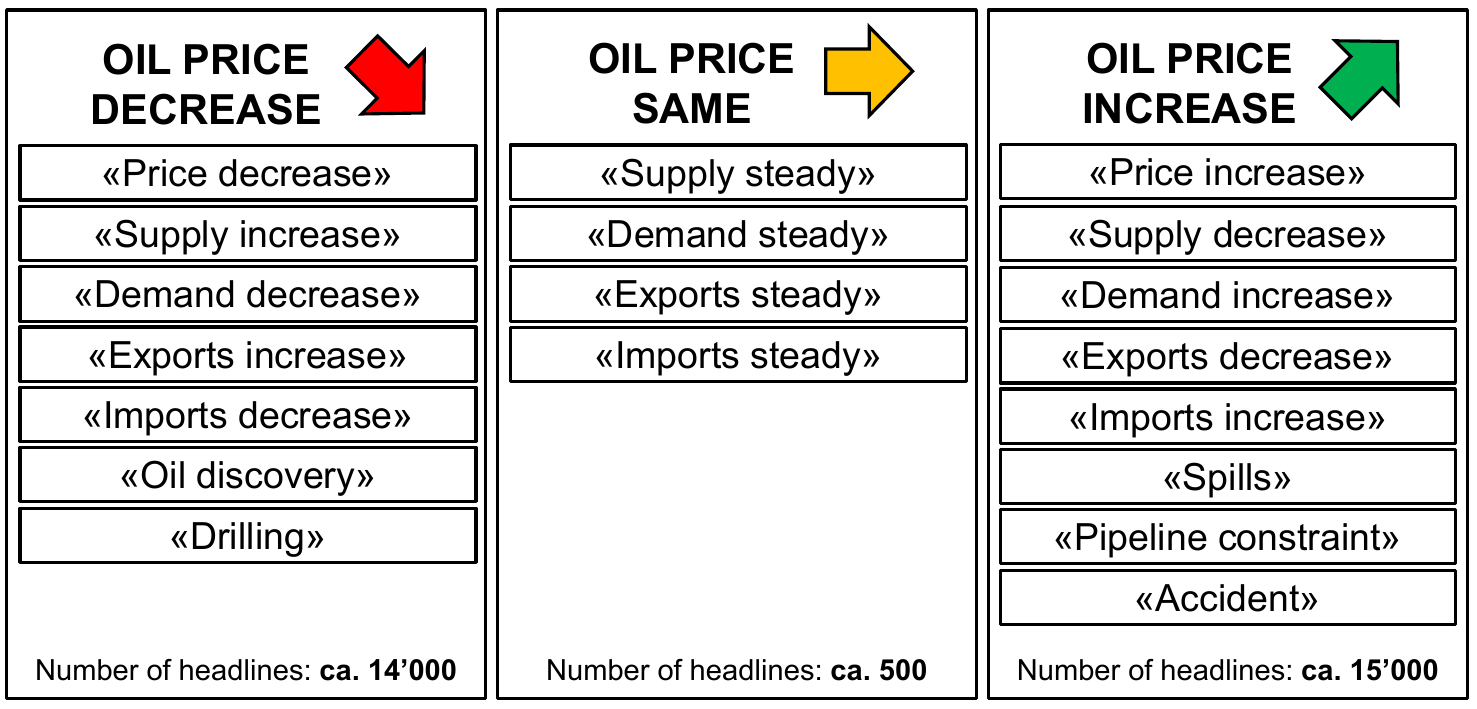}
\caption{Categorization of frequent recurring topics based on price theory~\cite{kaplan_conference_2023}.}
\label{fig:sd-dataset}
\end{figure}

\noindent Following this, the nature of the change, be it a rise, fall, or stagnation, was deduced by discerning the polarity of the headline, hinging on words indicative of these changes. A domain-specific silver standard emerged after evaluating these labels against the supply and demand price theory. This standard, represented in Figure~\ref{fig:sd-dataset}, was designed to classify headlines as either:

\begin{itemize}
\item \textbf{\emph{Price Decrease (score: $-1$):}} News detailing events like increased drilling activities, oil discoveries, augmented exports, or a mere surge in oil production are indicators of heightened supply. Additionally, a decrement in oil imports or consumption implies an oil surplus, leading to a price reduction.
\item \textbf{\emph{Price Increase (score: $+1$):}} Announcements of accidents, pipeline issues, oil spills, or a direct supply decline are suggestive of potential oil shortages. Moreover, a surge in demand or imports, and a decrease in exports, are indicative of possible oil scarcities, thus, a likely price increment.
\item \textbf{\emph{Stagnant Prices (score: $0$):}} A select few headlines that neither hint at shifts in supply, demand, imports, nor exports were ascribed a neutral score.
\end{itemize}

\begin{figure}
\centering
\includegraphics[width=\textwidth]{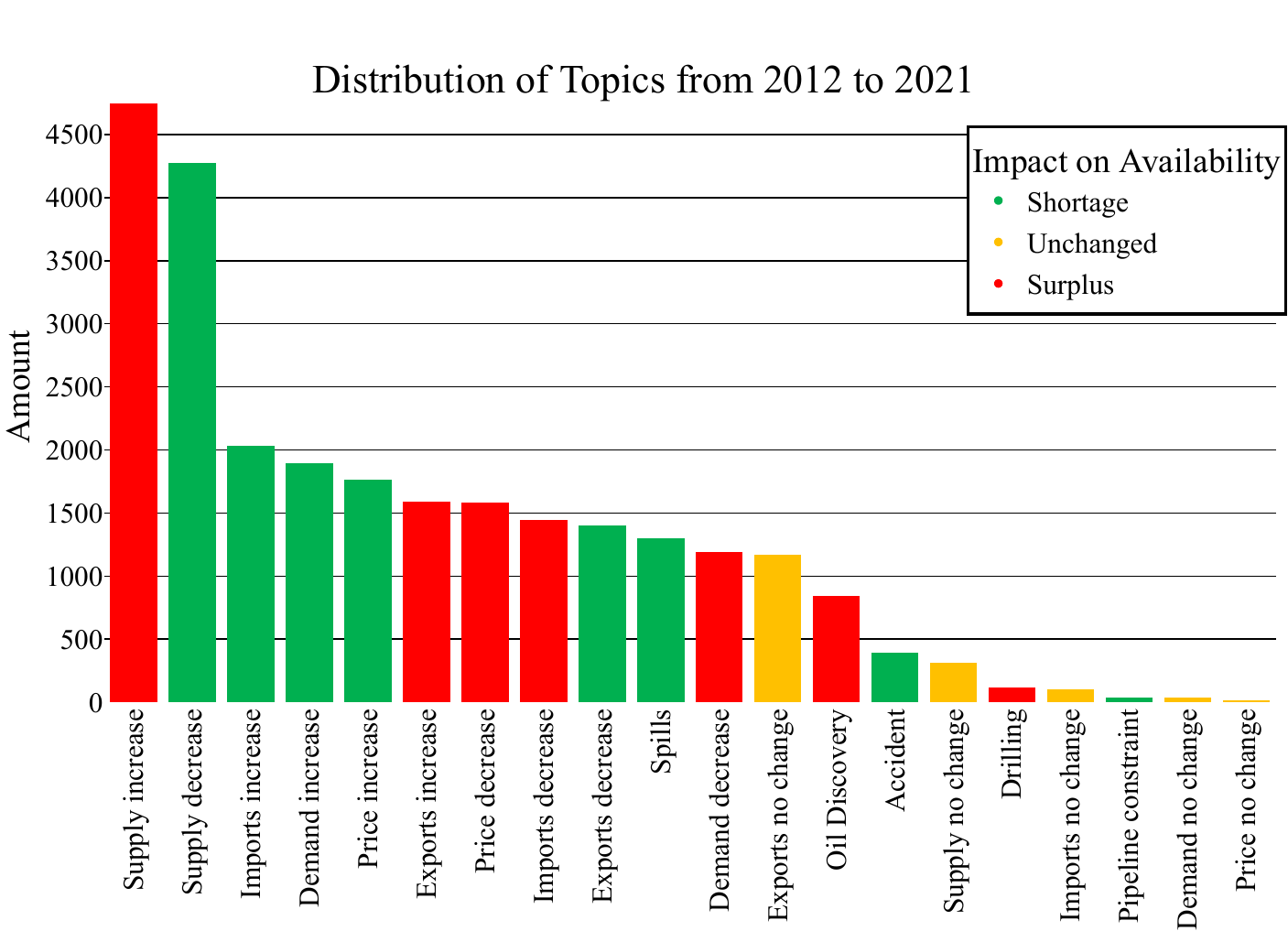}
\caption{Most recurring topics and their impact on the availability of crude oil.}
\label{fig:topic-histogram}
\end{figure}

\noindent This methodology enabled the computerized labeling of approximately 30,000 headlines with topics and their directional shifts. The generated output of these efforts can be seen in Figure \ref{fig:topic-histogram}, visualizing the top ten recurring topics.

\subsubsection{Fine-Tuning FinBERT}
The headlines, once labeled with domain-specific sentiment scores, were combined into the S\&D dataset. This dataset comprised around 14,000 negative, 500 neutral, and 15,000 positive headlines. The result was the emergence of the CrudeBERT classifier, as depicted in Figure~\ref{fig:process-crudebert}:
\begin{figure}[H]
\centering
\includegraphics[width=\textwidth]{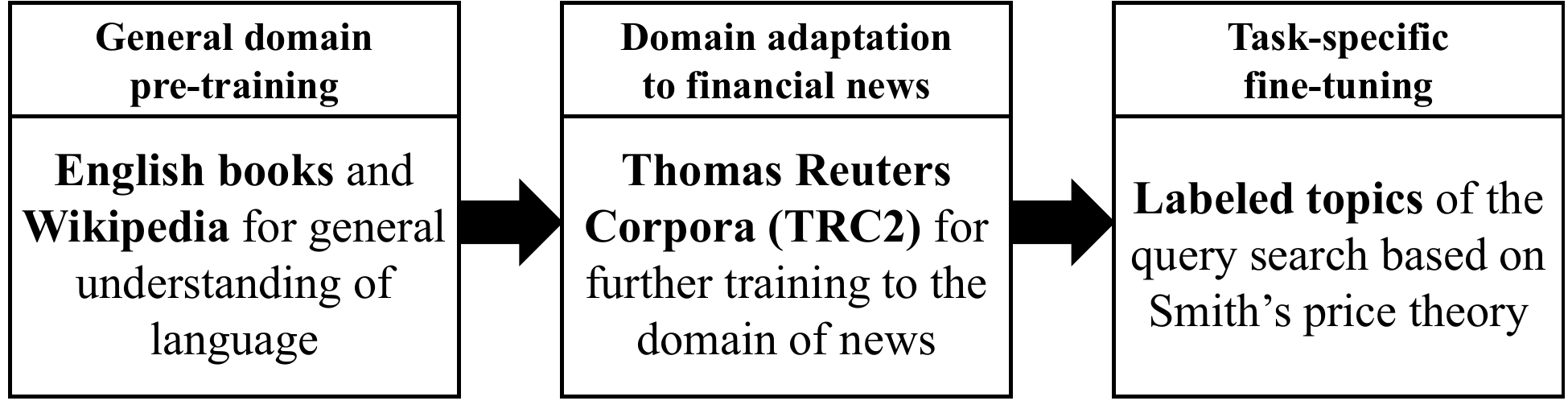}
\caption{Transformation from FinBERT to CrudeBERT through fine-tuning~\cite{kaplan_conference_2023}.}
\label{fig:process-crudebert}
\end{figure}

\begin{figure}[H]
\centering
\subfigure[FinBERT]{\includegraphics[width=.495\textwidth]{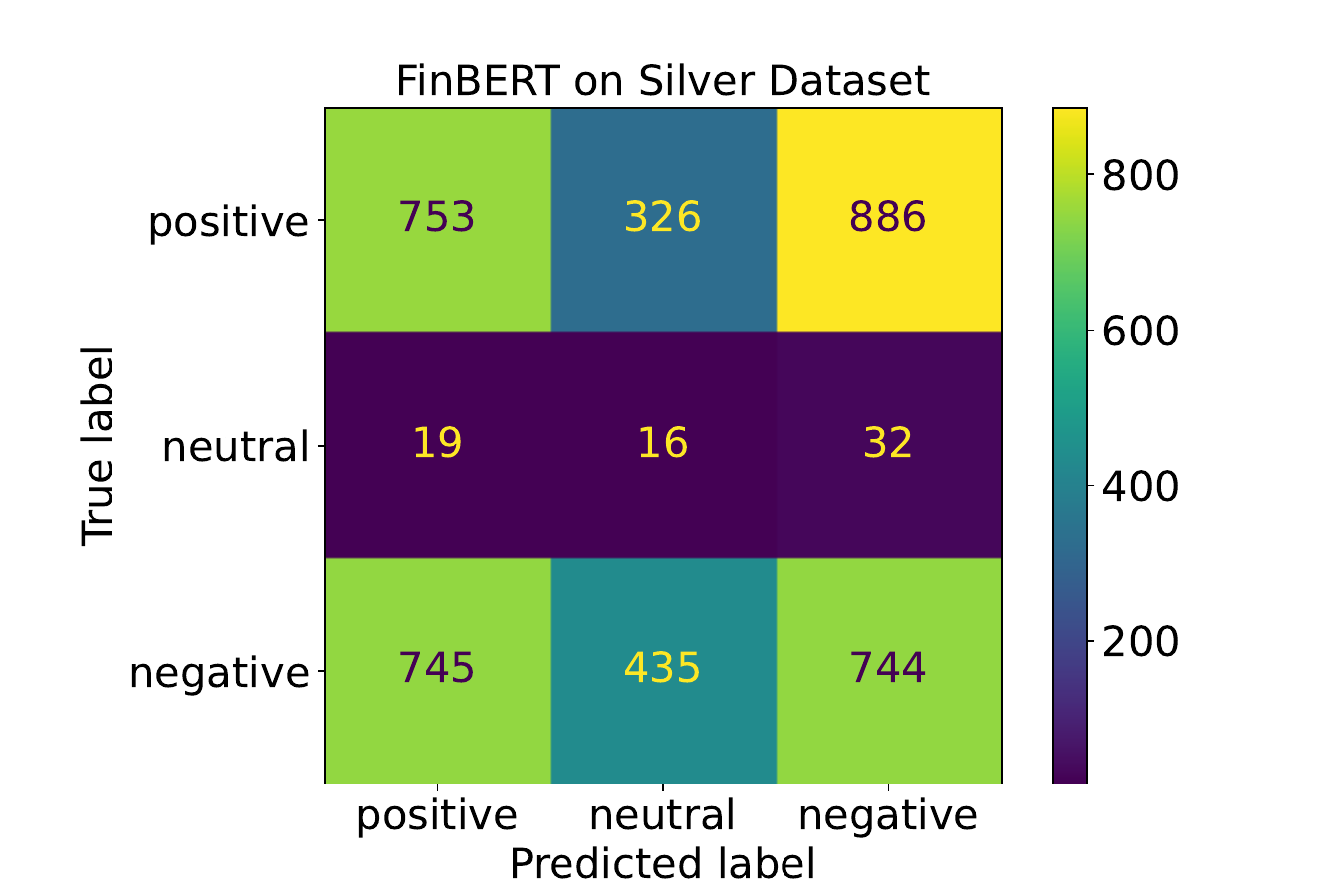}}
\subfigure[CrudeBERT]{\includegraphics[width=.495\textwidth]{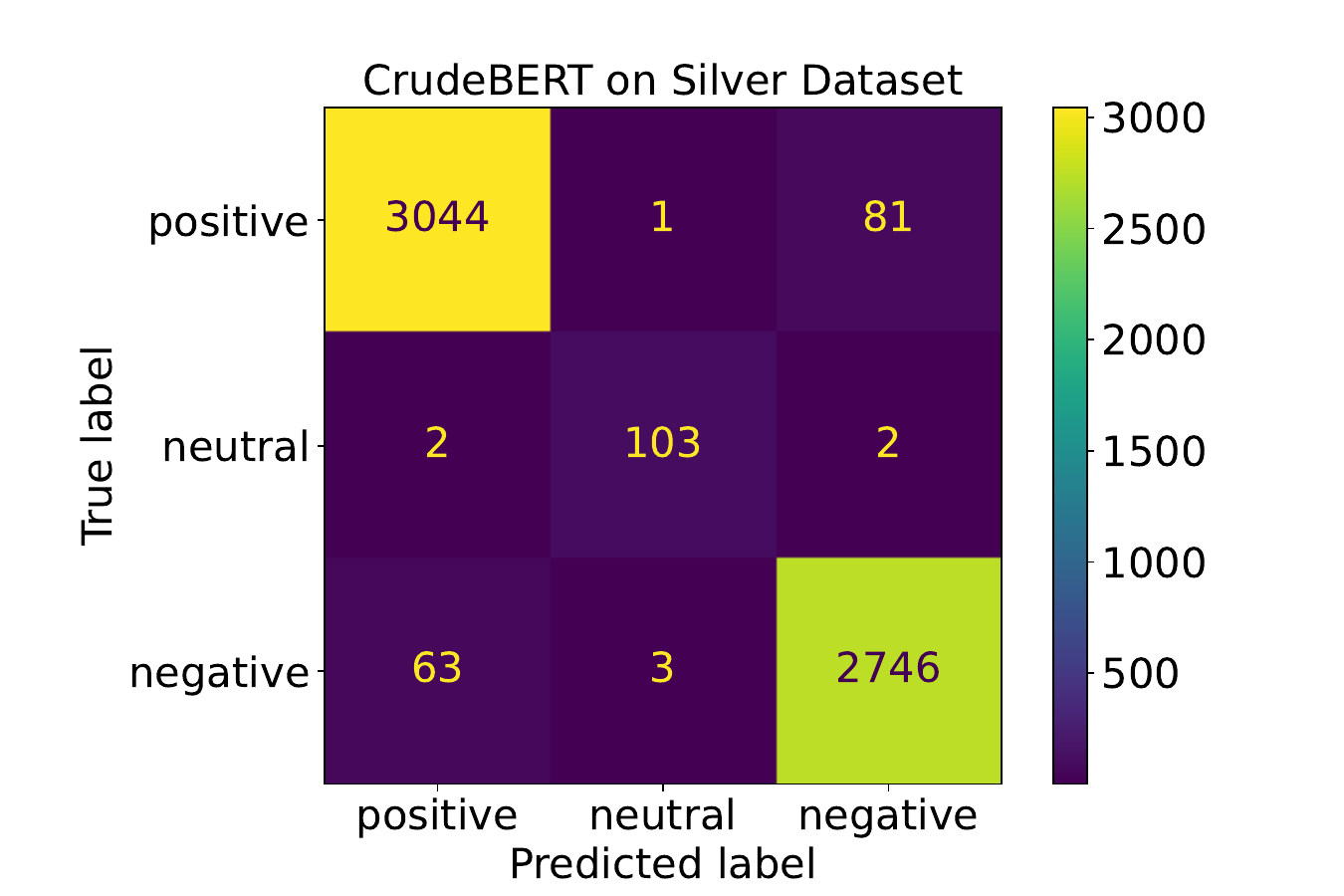}}
\caption{Confusion matrices of LMs before (left) and after fine-tuning (right)~\cite{kaplan_conference_2023}.} 
\label{fig:comparison-silver-dataset}
\end{figure}

\noindent The preliminary assessment of CrudeBERT on the silver standard dataset was remarkably high, especially considering the class imbalance. This dataset was divided into training (60\%), testing (20\%), and validation (20\%) portions, ensuring a uniform distribution of classes across these splits. Despite a comparatively sparse number of neutral headlines, their inclusion was deemed necessary. This ensured the model was introduced to domain-specific sentiment instances not situated at either extreme. Subsequently, FinBERT was fine-tuned using the test dataset. The classifier attained a weighted macro F1 score of 0.97. In contrast, FinBERT's performance was as expected substantially lower, with a weighted macro F1 score of 0.42 on the silver test dataset (Figure~\ref{fig:comparison-silver-dataset}).

\subsubsection{Prompt-Optimizing GPT}

Unlike FinBERT, GPT 3.5 was left unaltered owing to the high costs associated with such a large model. Instead, efforts were channeled into devising an optimal prompt, relying on the benchmark headlines from Table~\ref{tab:should-sentiment}. For this purpose, nine distinct prompts were composed and simulated to measure their effectiveness, where each prompt was directed to classify sentiments of the aforementioned headlines, marking them as 'Positive', 'Negative', or 'Neutral'. The simulations were:

\begin{itemize}
\small
\item Sim 1: No context
\item Sim 2: Context of oil prices
\item Sim 3: Context of oil availability
\item Sim 4: Context of supply-demand
\item Sim 5: Context of supply-demand + pragmatism
\item Sim 6: Context of supply-demand + 16 examples of each class
\item Sim 7: Context of supply-demand + 16 examples of each class + pragmatism
\item Sim 8: Context of supply-demand + assigned topics
\item Sim 9: Context of supply-demand + assigned topics + pragmatism
\end{itemize}

\noindent The full content of each prompt can be retrieved from the Appendix \ref{sec:appendix}.

\subsection{Evaluation of Fine-Tuning and Prompt Optimization Results}
\label{sec:evaluation-gpt}
\begin{figure}
\centering
\includegraphics[width=\textwidth]{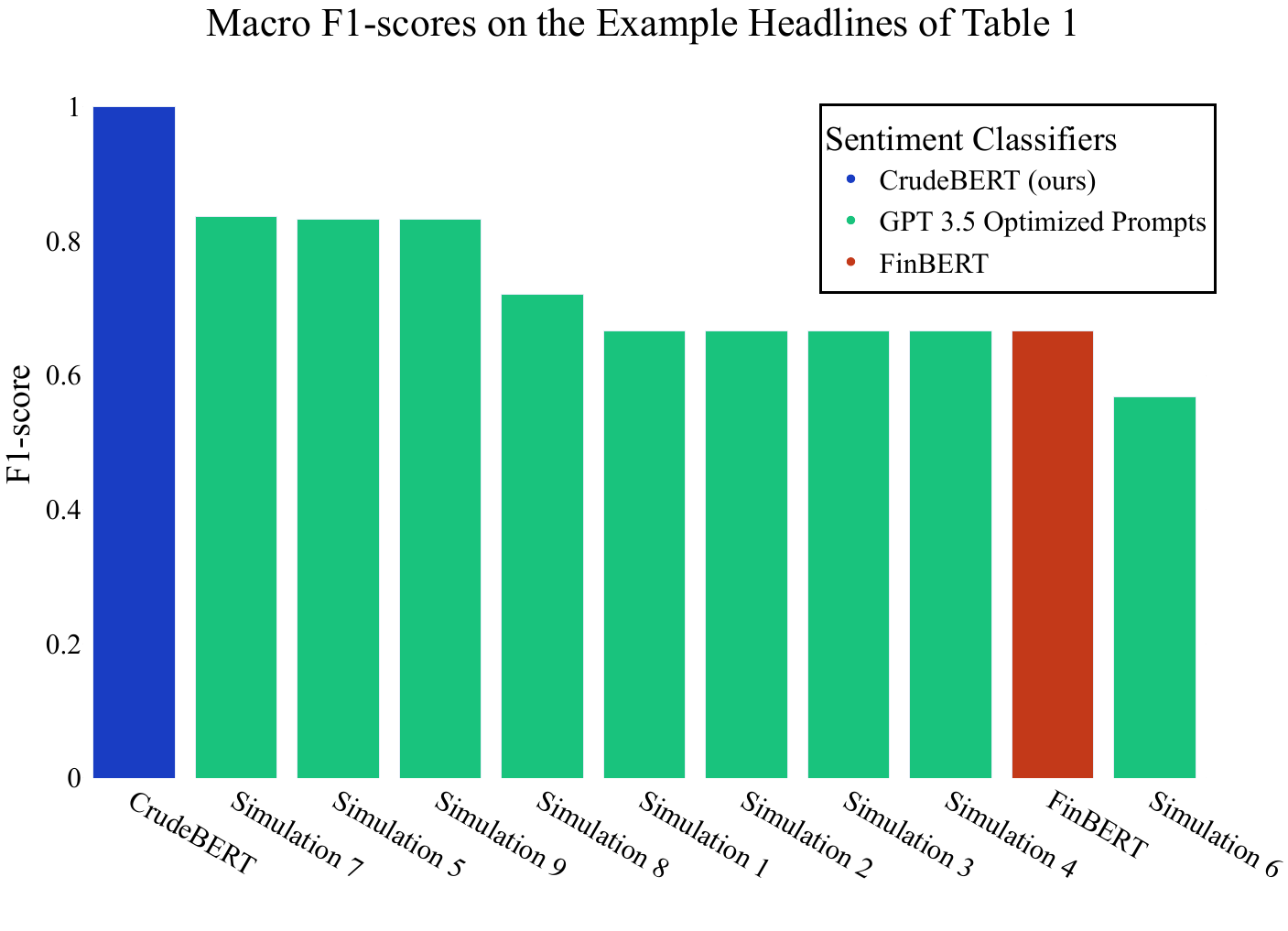}
\caption{Evaluation of fine-tuned LMs and prompt-optimized LLMs} 
\label{fig:gpt-prompt}
\end{figure}

A comparative study was undertaken using data presented in \ref{fig:gpt-prompt}. This involved the evaluation of different LMs including the financial LM, FinBERT, a fine-tuned LM, CrudeBERT, and an LLM, GPT, with each subjected to various prompt optimizations. Remarkably, CrudeBERT reached an F1 score of 1.0, setting it apart from the rest. From the insights gathered in \ref{fig:gpt-prompt}, Simulations 5, 7, and 9 displayed an enhanced performance for the LLM GPT. A notable macro F1 score of 0.84 was recorded, presenting a significant leap from the earlier 0.67 scores when the context was excluded from the prompt. A majority of these prompts efficiently categorized neutral sentiments. However, challenges emerged in sentiment interpretation, especially with headlines linked to accidents and injuries. Unexpectedly, in situations where sentiments were projected to be positive, perhaps due to predicted supply shortages, the models majorly categorized them as negative. To address this, Simulations 5, 7, and 9 were reconfigured to adopt a more discernible sentiment classification technique. 

As a result, Simulation 9 managed to classify sentiments as positive in the wake of accidents. The full result of each classifier is outlined in \ref{tab:preliminary-results} from the Appendix \ref{sec:appendix}.For the primary evaluation, the prompt from Simulation 9 was selected. This decision was based on its efficiency in using fewer tokens, while still achieving high performance when compared to other high-performing simulations. 

\subsection{Evaluation of Fine-Tuning and Prompt Optimization Results}
The following experiments leverage four different sentiment classifiers (FinBERT, CrudeBERT, RavenPack ESS, and GPT 3.5) including a random classifier as a baseline to assess the potential of analyzing headlines for predicting the direction of the next day's ($Return_{t+1}$) change in crude oil futures prices, using a two-class higher/lower price classification schema. The evaluation considers the period between 1 January 2012 and 1 April 2021 consisting of 3376 days' worth of data. We use precision, recall, and the F1 metric to assess the predictive potential of the evaluated classifiers. Table~\ref{tab:classification-report} presents the summarized evaluation results. On average, CrudeBERT outperforms FinBERT, RavenPack, GPT 3.5, and the baseline for binary classification.
Applying FinBERT without any customizations to the prediction task seems to be contra-productive since it yields worse results than the random baseline. However, Fine-tuning FinBERT with the presented domain adaptation method outlined in~\ref{sec:fine-tune} considerably improves the LMs performance. CrudeBERT's overall predictions also outperform those from prompt-optimized LLMs, specifically GPT 3.5. Compared to the proprietary sentiment classifier from Ravenpack, ESS, the margin by which CrudeBERT has an edge is narrower but remains consistent overall.

\begin{table}
  \centering
\caption{Classification Report of Different Sentiment Classifiers for Predicting Following Day WTI Oil Futures.}\label{tab:classification-report}
  \resizebox{\columnwidth}{!}{%
  \begin{tabular}{llllllll} 
    \toprule
   {\thead{Metric}} & {\thead{Category}} & {\thead{Random}} & {\thead{RavenPack}} & {\thead{FinBERT}} & {\thead{CrudeBERT}} & {\thead{GPT 3.5}}  \\
    \midrule
    Precision & Price down & 0.51   & 0.51  & 0.49 & 0.53 & 0.51\\
              & Price up   & 0.50   & 0.51  & 0.44 & 0.53 & 0.54\\
              & Macro      & 0.51   & 0.51  & 0.46 & 0.53 & 0.53\\
              \midrule
    Recall    & Price down & 0.51   & 0.47  & 0.85 & 0.53 & 0.84\\
              & Price up   & 0.50   & 0.55  & 0.11 & 0.52 & 0.19\\
              & Macro      & 0.51   & 0.51  & 0.48 & 0.53 & 0.51\\
              \midrule
    F1-Score  & Price down & 0.51   & 0.49  & 0.62 & 0.53 & 0.63\\
              & Price up   & 0.50   & 0.53  & 0.18 & 0.52 & 0.28\\
              & Macro      & 0.51   & 0.51  & 0.40 & 0.53 & 0.46\\
    \bottomrule
  \end{tabular}
  }
\end{table}

\noindent The performance differences between various models were evaluated by using Pearson's chi-square test, available in the SciPy~\cite{scipy_url} stats package. In this comparison, CrudeBERT displayed 1774 correct predictions against 1643 for FinBERT, a statistically significant difference at the 0.05 level. Similarly, when compared to RavenPack, CrudeBERT had 1774 correct predictions versus 1721 for RavenPack, significant at the 0.10 level. Lastly, CrudeBERT's 1774 correct predictions contrasted with GPT-3.5's 1739, but this disparity did not meet statistical significance at traditional thresholds.

\noindent However, it is essential to understand that while the Pearson chi-square test provides insights into the number of correct predictions, it might not capture the complete picture of a model's practical performance. This is especially true when certain models may show high precision but suffer in recall, which is not directly reflected in chi-square values. For a more holistic view, it's valuable to consider combined metrics like the macro F1 score, which harmoniously integrates both precision and recall of all classes. 

\begin{figure*}
\centering
\subfigure[RavenPack]{
\includegraphics[width=.47\textwidth]{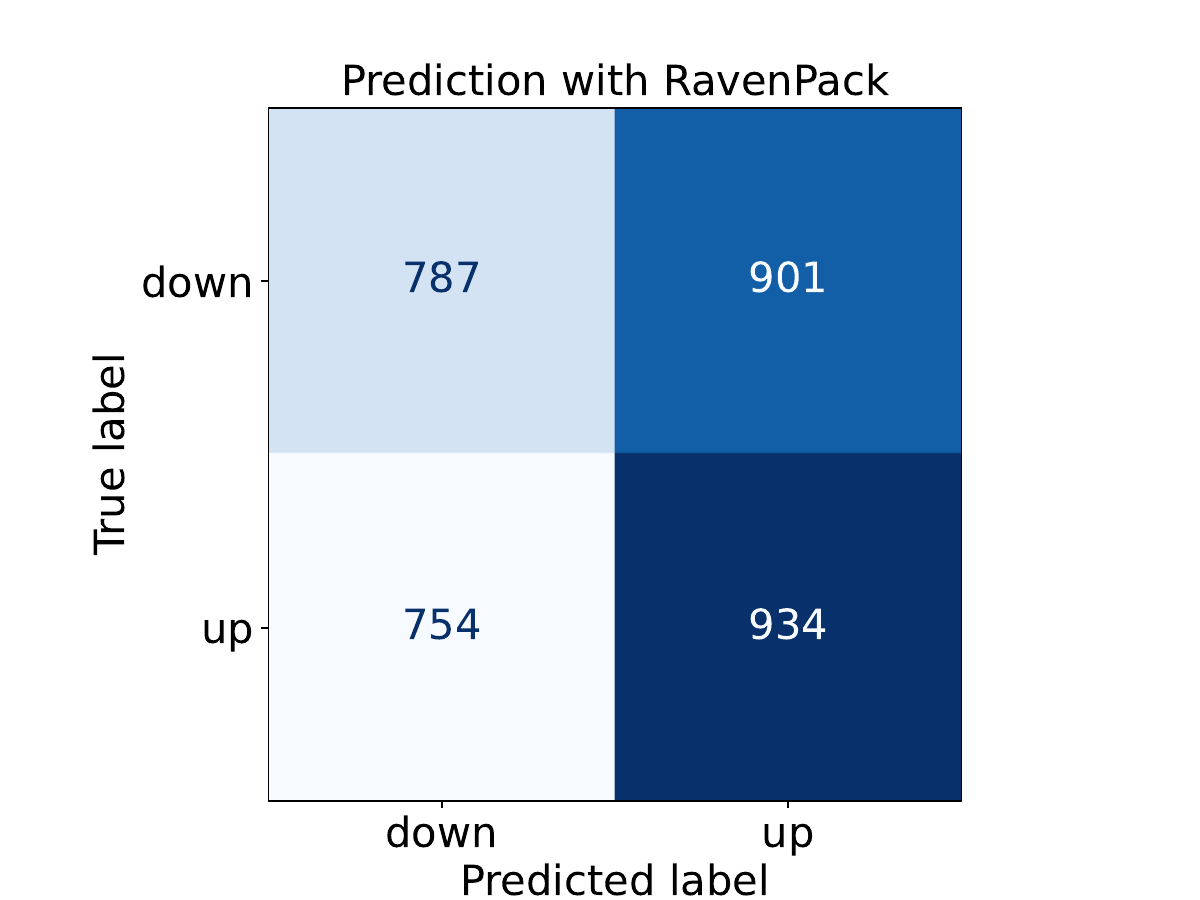}
}
\subfigure[FinBERT]{
\includegraphics[width=.47\textwidth]{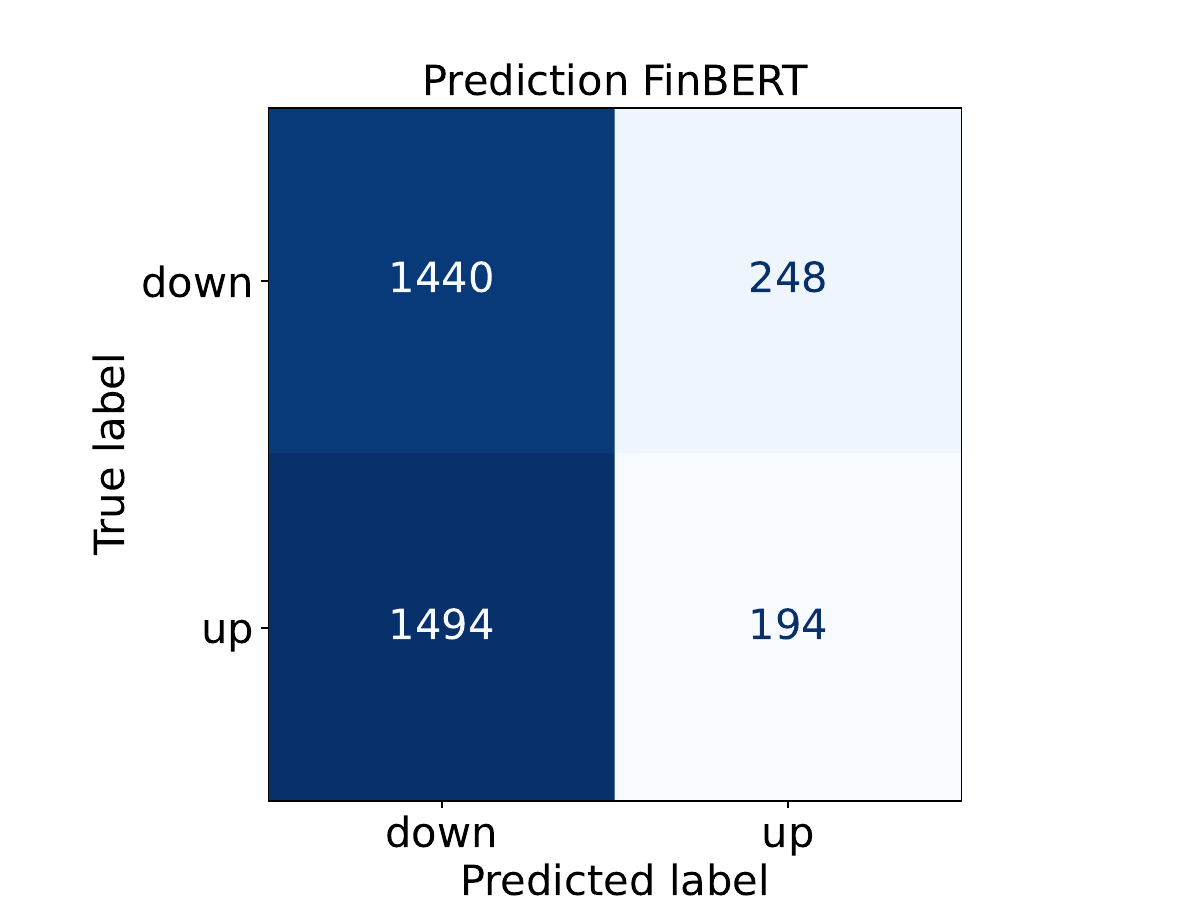}
}
\subfigure[CrudeBERT]{
\includegraphics[width=.47\textwidth]{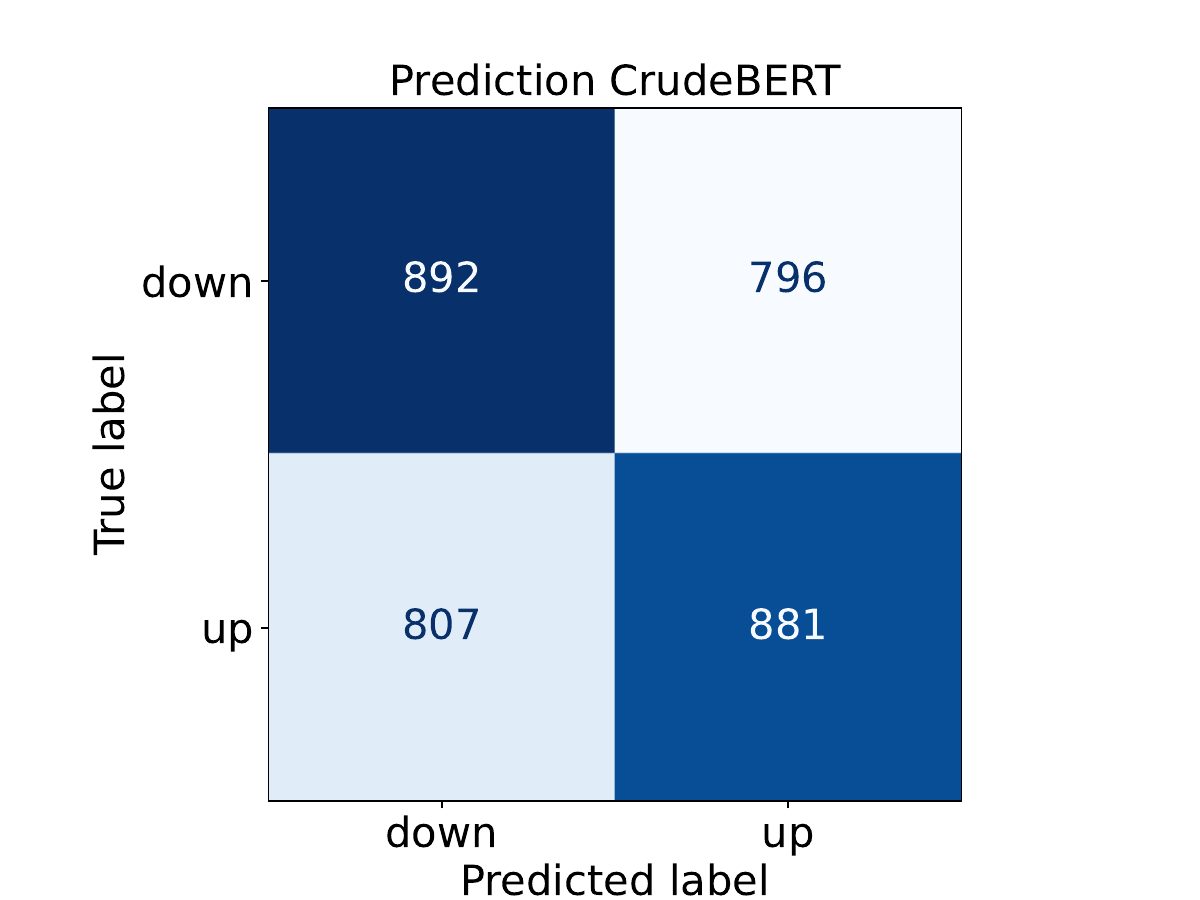}
}
\subfigure[GPT 3.5]{
\includegraphics[width=.47\textwidth]{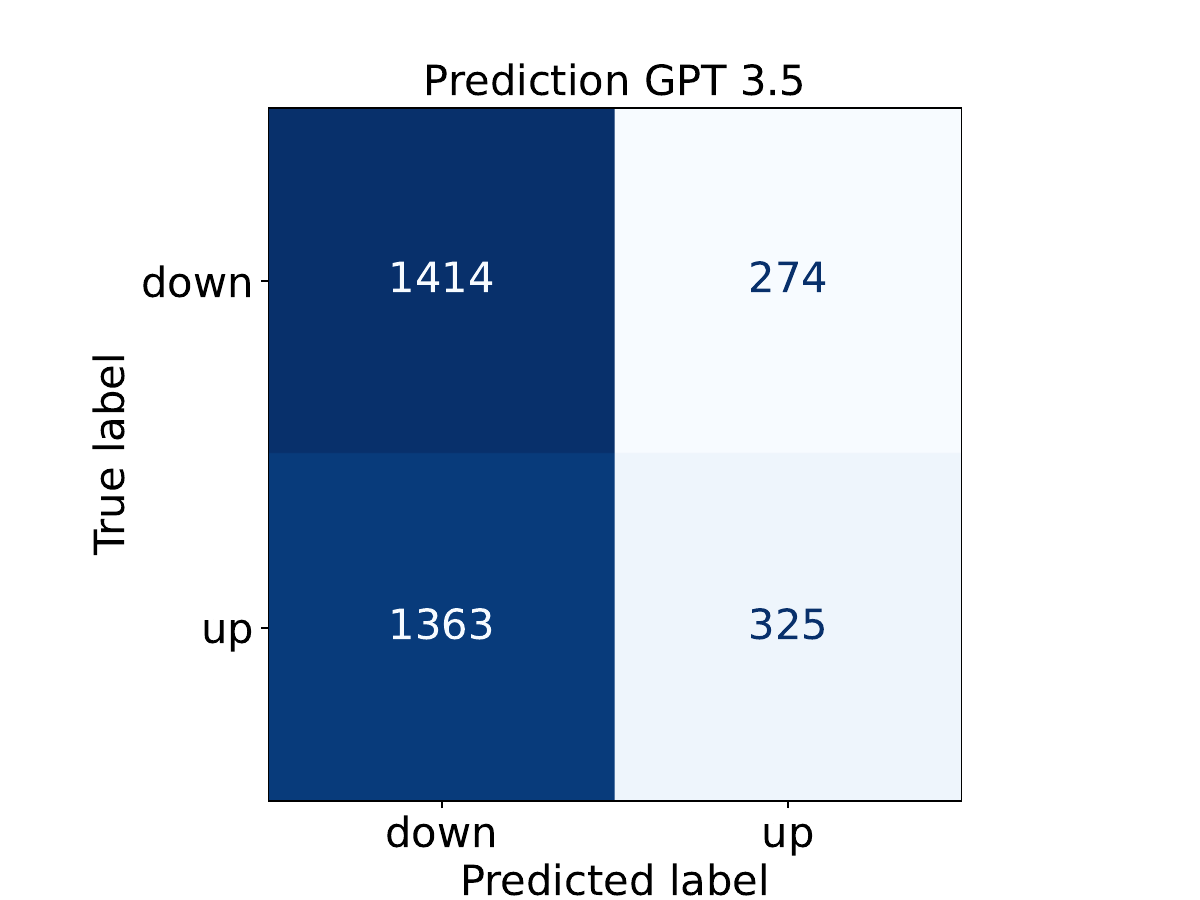}
}
\caption{Confusion matrices of the following day price predictions of four different sentiment classifiers.}
\label{fig:comparison-confusion-matrices}
\end{figure*}

In this context, Figure~\ref{fig:comparison-confusion-matrices} presents confusion matrices that compare the predicted label for each classifier with the following day's price changes of WTI crude oil futures ($Return_{t+1}$). This visual evaluation underscores the importance of comprehensive metric consideration, suggesting that while chi-square results offer one perspective, F1 scores offer a more encompassing view of model efficiency and reliability.

\begin{figure}
\centering
\includegraphics[width=\textwidth]{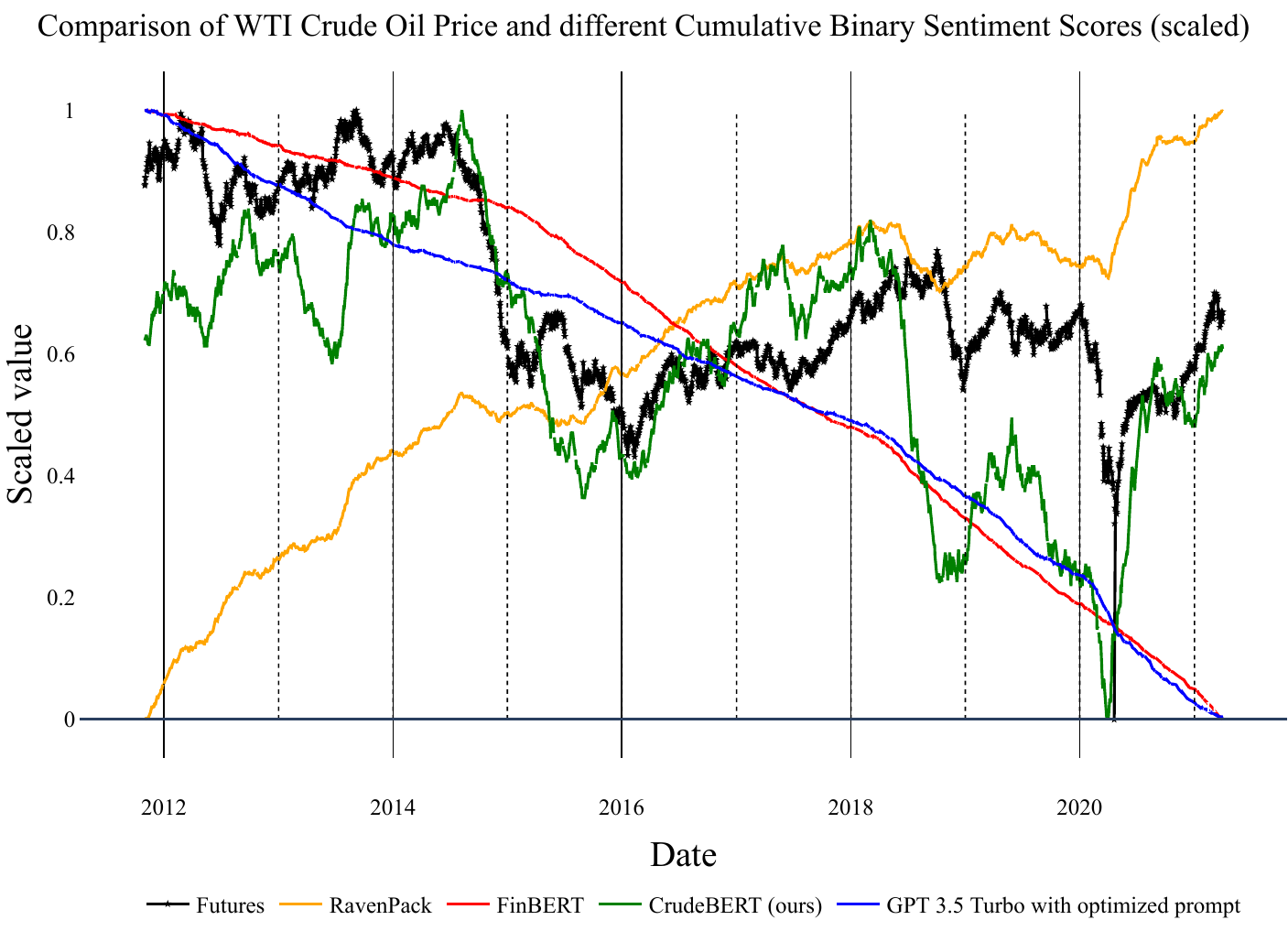}
\caption{Comparison of WTI oil prices with cumulative binary sentiment scores} 
\label{fig:cumulative-values}
\end{figure}

\noindent Furthermore, the qualitative comparison in Figure~\ref{fig:cumulative-values} further supports our initial intuition that FinBERT's lack of asset-specific knowledge of supply and demand seriously limits its suitability for prediction tasks. Consequently, it fails to track historical price movements compared to the fine-tuned CrudeBERT model and the commercial classifier of RavenPack. Similarly to the limitations observed with FinBERT, inconsistencies with GPT-3.5 were apparent. Despite the use of an optimized prompt, difficulties were observed in GPT-3.5's ability to accurately classify certain news topics outlined in Section~\ref{sec:evaluation-gpt}. For instance, consistent classification of articles regarding oil discoveries was not achieved. Additionally, inconsistencies were found in its handling of articles about oil imports. Such inconsistencies underline the challenges faced by models that might lack domain-specific knowledge. When compared qualitatively to the price curve the limitations of GPT-3.5 and FinBERT became more evident.

\noindent In conclusion, CrudeBERT remains the most effective sentiment classifier in this evaluation compared to the other proprietary (RavenPack ESS, GPT 3.5) and open-source (FinBERT) sentiment classifiers. The output of CrudeBERT could offer valuable insights into potential improvements and advancements in the realm of sentiment analysis for crude oil futures predictions.

\section{Outlook and Conclusion}
\label{sec:outlook}
The challenge of predicting market movements from the news was highlighted by the experiments described in Section~\ref{sec:evaluation}. Even advanced models like FinBERT, designed for the general financial domain, faced difficulties when attempting to classify the dynamics of a specific market such as crude oil without fine-tuning to the specific domain. Similarly, despite the multifaceted applications of LLMs, ranging from code generation to poem creation, the assessments of GPT 3.5 have underscored certain constraints. Specifically, these evaluations highlight the challenges LLMs face in extensive classification endeavors when dependent exclusively on an optimized prompt.
Moreover, the fine-tuning process of LLMs demands considerable computational and financial resources compared to that of LMs. For assignments with a narrower scope, such as straightforward classification, it might be both economical and efficient to adapt and use a smaller LM tailored for the specific NLP task at hand.

As a remedy, this paper introduced a method for refining LMs, in this case, FinBERT, using domain-specific news headlines. Through a frequency analysis, recurrent topics related to core market factors that affect supply and demand were identified. These topics were then used as search terms to filter and categorize headlines. For this purpose, a 'silver standard' dataset was created for a more reasonable sentiment classification based on the fundamental logic behind supply and demand, which was required in fine-tuning FinBERT into CrudeBERT. 

In evaluations covering nine years of crude oil futures, it was found that CrudeBERT performed better than both FinBERT, GPT 3.5, and a randomized baseline. Additionally, CrudeBERT showed competitive results against RavenPack's established sentiment model, although the differences were only noticeable at the 0.10 significance level. It is also noteworthy that news headlines alone rather than the whole article seem to be sufficient for providing insights into the likely direction of price changes.

Despite the presented improvements, CrudeBERT still has limitations and will be subject to further developments. For instance, concerns were raised regarding the creation process of the training data for fine-tuning. The current method based on keyword filtering may not effectively handle headlines that provide mixed and contradicting signals regarding supply-demand dynamics.

Lastly, research is needed to assess CrudeBERT's performance over longer periods and its adaptability across different economic conditions. One observation made was that headlines, rather than full articles, appeared to provide adequate information for resembling the market movement. Furthermore, the feasibility of extending the presented method to other commodity markets such as perishable (e.g., coffee beans), non-perishable (e.g., natural gas), precious (e.g., gold), and non-precious (e.g., iron ore) commodities, where pricing may be influenced by similar factors, need also to be investigated. 

Thus, to promote further research in asset-specific LMs, the CrudeBERT is made publicly available on the community-driven \textit{Hugging Face} platform~\cite{hf_url}.

\section*{Acknowledgement}
We would like to extend our gratitude to BRIDGE, the joint program of the Swiss National Science Foundation SNSF, and Innosuisse the Swiss Agency for Innovation Promotion and funding at the interface of basic research and science-based innovation~\cite{bridge_url}. In addition, we like to thank Prof. Dr. Hans Wernher van de Venn and the Institute of Mechatronic Systems at Zurich University of Applied Sciences for their generous support of this research. Lastly, we would like to thank Dr. Adrian M.P. Bra\c soveanu and Dr. Sahand Haji Ali Ahmad for their valuable input on suitable evaluations for the CrudeBERT language model.


\clearpage
\section*{Appendix}
\label{sec:appendix}

\subsection*{Unlabeled Test Dataset}
\label{sec:unlabeled-dataset}
\begin{mdframed}
\scriptsize\textbf{Unlabeled Test Dataset:}
\begin{enumerate}
\scriptsize
\item Major Explosion, Fire at Oil Refinery in Southeast Philadelphia
\item PETROLEOS confirms Gulf of Mexico oil platform accident
\item CASUALTIES FEARED AT OIL ACCIDENT NEAR IRANS BORDER
\item EIA Chief expects Global Oil Demand Growth 1 M B/D to 2011
\item Turkey Jan-Oct Crude Imports +98.5\% To 57.9M MT
\item China’s crude oil imports up 78.30\% in February 2019
\item Russia Energy Agency: Sees Oil Output Flat In 2005
\item Malaysia Oil Production Steady This Year At 700,000 B/D
\item ExxonMobil: Nigerian Oil Output Unaffected By Union Threat
\item Yukos July Oil Output Flat On Mo, 1.73M B/D - Prime-Tass
\item 2nd UPDATE: Mexico's Oil Output Unaffected By Hurricane
\item China CNPC '06 Domestic Crude Oil Output Flat
\item China February Crude Imports -16.0\% On Year
\item Turkey May Crude Imports down 11.0\% On Year
\item Japan June Crude Oil Imports decrease 10.9\% On Yr
\item Iran’s’ Feb Oil Exports +20.9\% On Mo at 1.56M B/D - Official
\item Apache announces large petroleum discovery in Philadelphia
\item Turkey finds oil near Syria, Iraq border
\end{enumerate}
\end{mdframed}

\subsection*{Labeled Training Dataset}
\label{sec:labeled-dataset}
\begin{mdframed}
\scriptsize\textbf{Labeled Training Dataset:}
\begin{enumerate}
\scriptsize
\item \textbf{Negative:} EIA: US Revised August Oil Demand -8.4\% Vs Yr Ago
\item \textbf{Negative:} Iraq Feb Oil Exports Rose by 7.5\% on Month to 2.536 Million B/D -- Sources
\item \textbf{Negative:} Global oil demand dropped 9 percent in 2020
\item \textbf{Negative:} OPEC expects global oil demand to drop 9.5\% in 2020
\item \textbf{Negative:} Russian Surgut 1st Quarter Crude Output Up 9.9\% On Year At 9.7 Mio MT
\item \textbf{Negative:} Colombia's Dec Oil Output Climbs 12\% On Year To 930,000 Bbls/Day
\item \textbf{Negative:} Ecuador Jan-July Crude Output 516,197 B/D; Up 36\% On Yr
\item \textbf{Negative:} Iraqi oil exports up 10 percent over past year
\item \textbf{Negative:} Africa Oil price target lowered to SEK 12 from SEK 14 at RBC Capital
\item \textbf{Negative:} MARKET TALK: Deutsche Bank Cuts '09 Oil Forecast 35\% To \$60
\item \textbf{Negative:} LukOil(OAO) Oil field discovery in Iran
\item \textbf{Negative:} Faroe Petroleum finds oil at Snilehorn in Norwegian Sea
\item \textbf{Negative:} Exxon Mobil discloses oil discovery offshore Guyana
\item \textbf{Negative:} TABLE-Germany's 2014 oil import bill down 10.5 percent
\item \textbf{Negative:} Japan's 2021 crude imports seen falling 5-10\% on year amid COVID-19
\item \textbf{Negative:} DJ China May Oil Product Imports 2.52 Mln Tons, Down 16\% on Year
\item \textbf{Neutral:} DJ Iraq To Export 1.25M B/D Oil In Nov, 1.5M B/D Jan - SOMO
\item \textbf{Neutral:} Angola June Oil Export Plan Has Five Hungo Cargoes - Trade
\item \textbf{Neutral:} Basra Oil Exports Unaffected By Iraq Pipeline Fire
\item \textbf{Neutral:} BRIEF-Enbridge revises estimate for July oil spill
\item \textbf{Neutral:} OPEC 'consensus' to hold oil output steady
\item \textbf{Neutral:} IRAQ KURDS SAY OIL PRODUCTION UNAFFECTED - STATEMENT
\item \textbf{Neutral:} Moody's Revises Outlook for Saudi Banks to Negative on Oil Slump
\item \textbf{Neutral:} Saudi calls Opec members to stick to oil output limits
\item \textbf{Neutral:} IEA '07 World Oil Demand Estimate Unchanged At 86.4M B/D
\item \textbf{Neutral:} US Apr Oil Demand Unch Vs Yr-Ago At 20.18M B/D -EIA
\item \textbf{Neutral:} IEA: 2001 Crude Demand Forecast Unchanged At 76.4M B/D
\item \textbf{Neutral:} India Mulls 40\% Import Tax On Crude,Refined Edible Oil-Report
\item \textbf{Neutral:} Limited refining expansion, high prices cloud China's 2021 crude import outlook
\item \textbf{Neutral:} Japan FY03 Crude Oil Imports Flat At 244.90M KL
\item \textbf{Neutral:} South Korea Aug Crude Imports At 72.8M Bbl, -3.8\% On Year
\item \textbf{Neutral:} Moody's Revises Outlook for Saudi Banks to Negative on Oil Slump
\item \textbf{Positive:} Eni says oil pipeline blast killed 12 in Niger Delta
\item \textbf{Positive:} Eni says oil pipeline blast killed 12 in Niger Delta
\item \textbf{Positive:} 12 injured after huge oil tanker blast in South Korea
\item \textbf{Positive:} China's 1H Crude Oil Imports +11.2\% On Yr At 81.54 Mln Tons
\item \textbf{Positive:} DJ S Korea's Mar Pete Pdts Exports -27\% On Yr, Imports +1.9\%
\item \textbf{Positive:} UPDATE 2-China March oil demand up 11 pct on year, but off peak
\item \textbf{Positive:} IEA: Oil demand to rise 14\% by 2035
\item \textbf{Positive:} Saudi Non Oil Exports Fall 10\% on Yr to SAR14.2 Bln in September
\item \textbf{Positive:} Advantage Oil price target raised to C\$3 from C\$2.25 at CIBC
\item \textbf{Positive:} China Petroleum Target Raised To HK\$12.40 Vs HK\$8.80 By Yuanta
\item \textbf{Positive:} Northern Oil Price Target Raised to \$2.5/Share From \$2.0 by Imperial Capital
\item \textbf{Positive:} Kazakhstan to Cut Oil Output by 20,000 Bpd - Energy Minister
\item \textbf{Positive:} Alaska Oil Output Drops 11\% as North Slope Production Declines
\item \textbf{Positive:} Sinopec Corp: 2016 Crude-Oil Production Down 13\% on Year
\item \textbf{Positive:} Reliance May crude import up 23 pct y/y - trade
\item \textbf{Positive:} Russia '00 Crude Exports Seen Dn 7.5\% At 126-128 Mln MT
\end{enumerate}
\end{mdframed}

\subsection*{Simulation Prompts}
\label{sec:simulation-prompts}

\begin{mdframed}[frametitle={\scriptsize\textbf{Simulation 1 - No context}}]
\scriptsize
Classify the sentiment of the following headlines as either 'Positive', 'Negative', or 'Neutral'. 
Return only the ID and your classification as a dict named dict\_sim1 (do not provide any code or explanation, just the dict).
\\
\\
\newcommand{\datasetplaceholder}{\fbox{ $\rightarrow$ Unlabeled Test Dataset}}
\datasetplaceholder
\end{mdframed}

\begin{mdframed}[frametitle={\scriptsize\textbf{Simulation 2 - Context about oil prices}}]
\scriptsize
Classify the following headlines as either 'Positive', 'Negative', or 'Neutral' with regard to their impact on crude oil prices. Return only the ID and your classification as a dict named dict\_sim2 (do not provide any code or explanation, just the dict).
\\
\\
\newcommand{\datasetplaceholder}{\fbox{ $\rightarrow$ Unlabeled Test Dataset}}
\datasetplaceholder
\end{mdframed}

\begin{mdframed}[frametitle={\scriptsize\textbf{Simulation 3 - Context about oil availability}}]
\scriptsize
Classify the following headlines as either 'Positive', 'Negative', or 'Neutral' with regard to their impact on the availability of crude oil. Return only the ID and your classification as a dict named dict\_sim3 (do not provide any code or explanation, just the dict).
\\
\\
\newcommand{\datasetplaceholder}{\fbox{ $\rightarrow$ Unlabeled Test Dataset}}
\datasetplaceholder
\end{mdframed}

\begin{mdframed}[frametitle={\scriptsize\textbf{Simulation 4 - Context about supply and demand}}]
\scriptsize
Classify the following headlines as either 'Positive', 'Negative', or 'Neutral' with regard to their impact on crude oil prices based on the market theory of supply and demand. Return only the ID and your classification as a dict named dict\_sim4 (do not provide any code or explanation, just the dict).
\\
\\
\newcommand{\datasetplaceholder}{\fbox{ $\rightarrow$ Unlabeled Test Dataset}}
\datasetplaceholder
\end{mdframed}

\begin{mdframed}[frametitle={\scriptsize\textbf{Simulation 5 - Context about supply and demand + Pragmatism}}]
\scriptsize
Classify the following headlines as either 'Positive', 'Negative', or 'Neutral' with regard to their impact on crude oil prices based on the market theory of supply and demand.
Remain pragmatic (e.g. shortage-causing topics such as explosions, injuries, spills, etc. have a positive and surplus-causing discovery, drills, etc. have a negative impact). Return only the ID and your classification as a dict named dict\_sim5 (do not provide any code or explanation, just the dict).
\\
\\
\newcommand{\datasetplaceholder}{\fbox{ $\rightarrow$ Unlabeled Test Dataset}}
\datasetplaceholder
\end{mdframed}

\begin{mdframed}[frametitle={\scriptsize\textbf{Simulation 6 - Context about supply and demand + training data}}]
\scriptsize
Based on the training data, classify the following headlines as either 'Positive', 'Negative', or 'Neutral' with regard to their impact on crude oil prices based on the market theory of supply and demand. Return only the ID and your classification as a dict named \texttt{dict\_sim6} (do not provide any code or explanation, just the dict).
\vspace*{1em}
\newcommand{\datasetplaceholderA}{\fbox{ $\rightarrow$ Labeled Training Dataset}}
\newcommand{\datasetplaceholderB}{\fbox{ $\rightarrow$ Unlabeled Test Dataset}}

\datasetplaceholderA

\vspace*{1em}
\datasetplaceholderB
\end{mdframed}

\begin{mdframed}[frametitle={\scriptsize\textbf{Simulation 7 - Context about supply and demand + training data + Pragmatism}}]
\scriptsize
Based on the training data, classify the following headlines as either 'Positive', 'Negative', or 'Neutral' with regard to their impact on crude oil prices based on the market theory of supply and demand. Remain pragmatic (e.g. shortage-causing topics such as explosions, injuries, spills, etc. have a positive and surplus-causing discovery, drills, etc. have a negative impact). Return only the ID and your classification as a dict named dict\_sim7 (do not provide any code or explanation, just the dict).
\vspace*{1em}
\newcommand{\datasetplaceholderA}{\fbox{ $\rightarrow$ Labeled Training Dataset}}
\newcommand{\datasetplaceholderB}{\fbox{ $\rightarrow$ Unlabeled Test Dataset}}

\datasetplaceholderA

\vspace*{1em}
\datasetplaceholderB
\end{mdframed}

\begin{mdframed}[frametitle={\scriptsize\textbf{Simulation 8 - Context about supply and demand + assigned topics}}]
\scriptsize
Identify the topic of the following headlines and assign their impact on the crude oil prices as either 'Positive', 'Negative', or 'Neutral' and use the topic\_impact as a guide. Return only the ID and your classification as a dict named dict\_sim8 (do not provide any code or explanation, just the dict).

Negative impacts include price decreases and surpluses caused by supply increases, demand decreases, rising exports, falling imports, oil discoveries, and drilling.
Neutral impacts involve steady supply, stable demand, and consistent export and import levels.
Positive impacts include price increases and shortages caused by supply decreases, demand growth, reduced exports, increased imports, spills, pipeline constraints, and accidents.
\\
\\
\newcommand{\datasetplaceholder}{\fbox{ $\rightarrow$ Unlabeled Test Dataset}}
\datasetplaceholder
\end{mdframed}

\begin {mdframed}[frametitle={\scriptsize\textbf{Simulation 9 - Context about supply and demand + assigned topics + Pragmatism}}]

Identify the topic of the following headlines and assign their impact on the crude oil prices as either 'Positive', 'Negative', or 'Neutral' and use the information below as a guide. Remain pragmatic (e.g. shortage-causing topics such as explosions, injuries, spills, etc. have a positive and surplus-causing discovery, drills, etc. have a negative impact). Return only the ID and your classification as a dict named dict\_sim9 (do not provide any code or explanation, just the dict).
\\
Negative impacts include price decreases and surpluses caused by supply increases, demand decreases, rising exports, falling imports, oil discoveries, and drilling.
\\
Neutral impacts involve steady supply, stable demand, and consistent export and import levels.
\\Positive impacts include price increases and shortages caused by supply decreases, demand growth, reduced exports, increased imports, spills, pipeline constraints, and accidents.
\\
\\
\newcommand{\datasetplaceholder}{\fbox{ $\rightarrow$ Unlabeled Test Dataset}}
\datasetplaceholder
\end{mdframed}

\subsection*{Comparative Results of the Preliminary Evaluation}
\label{sec:initial-evaluation}
\vspace{-0.8cm}
\begin{table}[h!]
\centering
\begin{tabularx}{\textwidth}{c*{12}{X}}
\toprule
\textbf{Key} & \textbf{True} & \textbf{Sim1} & \textbf{Sim2} & \textbf{Sim3} & \textbf{Sim4} & \textbf{Sim5} & \textbf{Sim6} & \textbf{Sim7} & \textbf{Sim8} & \textbf{Sim9} & \textbf{FB} & \textbf{CB} \\
\midrule
\rowcolor{grayrow}
1 & P & N & N & N & N & P & N & P & N & P & N & P \\
2 & P & N & N & N & N & P & N & P & N & P & N & P \\
\rowcolor{grayrow}
3 & P & N & N & N & N & P & N & P & N & P & N & P \\
4 & P & P & P & P & P & P & P & P & P & P & P & P \\
\rowcolor{grayrow}
5 & P & P & P & P & P & N & N & N & P & N & P & P \\
6 & P & P & P & P & P & N & P & P & P & P & P & P \\
\rowcolor{grayrow}
7 & Ne & Ne & Ne & Ne & Ne & Ne & Ne & P & Ne & Ne & N & Ne \\
8 & Ne & Ne & Ne & Ne & Ne & Ne & P & Ne & Ne & Ne & P & Ne \\
\rowcolor{grayrow}
9 & Ne & Ne & Ne & Ne & Ne & Ne & Ne & Ne & Ne & Ne & N & Ne \\
10 & Ne & Ne & Ne & Ne & Ne & Ne & Ne & Ne & Ne & Ne & N & Ne \\
\rowcolor{grayrow}
11 & Ne & Ne & Ne & Ne & Ne & Ne & Ne & Ne & Ne & Ne & N & Ne \\
12 & Ne & Ne & Ne & Ne & Ne & Ne & Ne & Ne & Ne & Ne & Ne & Ne \\
\rowcolor{grayrow}
13 & N & N & N & N & N & N & N & N & N & N & P & N \\
14 & N & N & N & N & N & N & N & N & N & N & N & N \\
\rowcolor{grayrow}
15 & N & N & N & N & N & N & N & N & N & N & N & N \\
16 & N & P & P & P & P & P & P & P & N & P & Ne & N \\
\rowcolor{grayrow}
17 & N & P & P & P & P & N & P & N & P & P & Ne & N \\
18 & N & P & P & P & P & N & P & N & P & N & Ne & N \\
\bottomrule
\end{tabularx}
\caption{Results of GPT simulations, CrudeBERT (CB), and FinBERT(FB), where Positive = P, Neutral = Ne, and Negative = N}
\label{tab:preliminary-results}
\end{table}

\end{document}